\documentclass[a4paper,11pt]{article}
\pdfoutput=1

\usepackage{jcappub}
\usepackage{amsmath,amsfonts,amssymb,mathrsfs,graphicx,color,longtable,bm}
\usepackage{hyperref}
\usepackage{cancel}
\usepackage{graphicx}
\usepackage{array}
\usepackage{cases} 
\usepackage[normalem]{ulem}
\usepackage{enumitem}
\usepackage[explicit]{titlesec}
\usepackage[utf8]{inputenc}
\usepackage[normalem]{ulem}
\usepackage{cleveref}
\usepackage{booktabs}
\usepackage{xcolor}
\usepackage{arydshln}
\usepackage{multirow}
\usepackage{bm}
\usepackage{ulem}

\hypersetup{
     colorlinks   = true,
     citecolor    = red,
     urlcolor     = blue,
     linkcolor    = blue
}

\makeatletter
\gdef\@fpheader{}
\makeatother

\usepackage{physics}
\usepackage{lipsum}
\usepackage{subfigure}
\usepackage{verbatim}
\usepackage{cancel}
\usepackage{tikz}
\usetikzlibrary{mindmap, tikzmark, calc, decorations.pathreplacing}

\newcommand{\noleft}{\left.\kern-\nulldelimiterspace}
\newcommand{\noright}{\right.\kern-\nulldelimiterspace}
\newcommand{\round}[1]{\left( #1 \right)}
\newcommand{\sround}[1]{\left[ #1 \right]}

\newcommand{\PR}{P_\zeta}
\newcommand{\As}{A_\mathrm{s}}
\newcommand{\ns}{n_\mathrm{s}}
\newcommand{\rs}{r_\mathrm{s}}

\title{Signals from the early Universe: a comprehensive search for primordial features in Planck CMB datasets}

\author[a,b]{Antonio Raffaelli,}
\author[a,b,c,d]{Mario Ballardini,}
\author[e,b]{Nicola Barbieri}

\affiliation[a]{Dipartimento di Fisica e Scienze della Terra, Universit\`a degli Studi di Ferrara, via Giuseppe Saragat 1, 44122 Ferrara, Italy}
\affiliation[b]{INFN, Sezione di Ferrara, via Giuseppe Saragat 1, 44122 Ferrara, Italy}
\affiliation[c]{INAF-Osservatorio di Astrofisica e Scienza dello Spazio di Bologna, Via Piero Gobetti 93/3, 40129 Bologna, Italy}
\affiliation[d]{Department of Physics \& Astronomy, University of the Western Cape, Cape Town 7535, South Africa}
\affiliation[e]{Instituto de Física Corpuscular (IFIC), CSIC‐Universitat de València, Parc Científic UV, c/ Catedrático José Beltrán, 2, E-46980 Paterna (València), Spain}

\emailAdd{antonio.raffaelli@unife.it}
\emailAdd{mario.ballardini@unife.it}
\emailAdd{barbieri@ific.uv.es}

\abstract{We investigate the presence of primordial oscillatory features in measurements of cosmic microwave background (CMB) anisotropies through a systematic comparison of phenomenological templates. Building upon previous searches for primordial features using \textit{Planck} data, we compare the full PR3 legacy release with the PR4 (\texttt{NPIPE}) processing to assess how the results depend on the choice of CMB maps and likelihood framework. To maximise our sensitivity to rapidly varying oscillatory signals, we employ unbinned likelihoods.

We find that several previously reported indications of oscillatory structure persist across different analyses, although none attains global statistical significance. Furthermore, some anomalies reported in earlier studies are substantially reduced when updated to the new versions of the \texttt{CamSpec} likelihood using \textit{Planck} PR4 products.
For all templates considered, we identify a small number of frequencies in the range $\omega \sim 10$--$100$ that improve the fit to the CMB data by up to $\Delta\chi^2 \simeq -10$ to $-15$ relative to the featureless reference model. However, this improvement is not supported by a Bayesian model comparison, as the Bayes factor consistently does not favour the extended models. The inclusion of three or four additional parameters can reduces the overall predictability of the feature models and leads to an Occam penalty. Finally, after properly accounting for the look-elsewhere effect, the significance of the preferred frequencies is reduced, corresponding to a global statistical significance of at most $2.6\sigma$.

Finally, we present forecasts for forthcoming CMB experiments, highlighting the decisive role of next-generation polarisation measurements in distinguishing genuine primordial oscillations from statistical fluctuations and modelling systematics.
The upper bounds or uncertainties on the feature amplitudes, expected from the combination of Simons Observatory and LiteBIRD, improve by more than one order of magnitude.}

\begin{document}

\maketitle

\section{Introduction}
The inflationary paradigm has emerged as the leading framework for describing the early Universe, successfully addressing the horizon and flatness problems and providing a causal mechanism for the generation of primordial seed perturbations~\cite{Starobinsky:1980te,Guth:1980zm,Linde:1981mu,Linde:1983gd,Mukhanov:1981xt}. Current cosmological observations, most notably those from the \textit{Planck} satellite, are in excellent agreement with the predictions of the standard $\Lambda$ Cold Dark Matter (CDM) cosmological model~\cite{Planck:2018vyg,Akrami:2018odb,Planck:2019izv}. Within this framework, the primordial power spectrum (PPS) of curvature perturbations is parametrised as a nearly scale-invariant power law, $\PR(k) = \As(k/k_*)^{\ns-1}$. The precise measurement of the scalar spectral index, $\ns \sim 0.965$~\cite{Planck:2018vyg}, confirms a departure from exact scale invariance, in agreement with the predictions of the simplest single-field slow-roll inflationary scenarios, with no additional deviations detected over the range $0.005 \lesssim k \, \mathrm{Mpc} \lesssim 0.25$ once model-independent reconstructions of the PPS are taken into account~\cite{Akrami:2018odb,Handley:2019fll,Raffaelli:2025kew}.

However, a featureless power-law PPS remains only the minimal description consistent with current cosmological data. From a theoretical perspective, the physics of the early Universe is expected to be considerably richer. A broad landscape of inflationary scenarios predicts departures from simple slow-roll dynamics, leaving distinctive imprints on the PPS, commonly referred to as \textit{primordial features}; see~\cite{Chluba:2015bqa,2022arXiv220308128A} for reviews. Such features may arise from a variety of physical mechanisms, including sharp steps or kinks in the inflationary potential~\cite{Starobinsky:1992ts,Adams:2001vc,Chen:2006xjb}, particle-production events~\cite{Chung:1999ve,Romano:2008rr,Barnaby:2009mc}, or oscillations in the background evolution induced by massive fields~\cite{Chen:2008wn,Flauger:2009ab,Flauger:2010ja,Chen:2010bka,Achucarro:2010da,Chen:2011zf}. In general, these mechanisms generate oscillatory patterns in the PPS, either linear or logarithmic in the wavenumber $k$, which are projected onto the CMB anisotropies as intricate oscillatory signatures across a broad range of angular scales.

Over the past decade, extensive searches for such signals have been carried out using CMB observations, particularly \textit{Planck} data~\cite{Adams:2001vc,Peiris:2003ff,Mukherjee:2003ag,Covi:2006ci,Hamann:2007pa,Meerburg:2011gd,Planck:2013jfk,Planck:2013wtn,Meerburg:2013dla,Benetti:2013cja,Miranda:2013wxa,Easther:2013kla,Chen:2014joa,Achucarro:2014msa,Hazra:2014goa,Hazra:2014jwa,Hu:2014hra,Fergusson:2014tza,Planck:2015zfm,Ade:2015lrj,Gruppuso:2015zia,Gruppuso:2015xqa,Hazra:2016fkm,Torrado:2016sls,Gruppuso:2017nap,Akrami:2018odb,Zeng:2018ufm,Planck:2019izv,Canas-Herrera:2020mme,Braglia:2021ckn,Braglia:2021sun,Braglia:2021rej,Naik:2022mxn,Hamann:2021eyw,Antony:2024vrx,2026arXiv260413547H}. Although no signal has reached the threshold required for a detection, several anomalies and marginal hints have been reported in both the temperature and polarisation spectra. Revisiting these hints is therefore essential. Despite the growing emphasis on large-scale structure measurements~\cite{Huang:2012mr,Chen:2016vvw,Ballardini:2016hpi,Ballardini:2017qwq,Beutler:2019ojk,Ballardini:2019tuc,Ballardini:2022wzu,Ballardini:2022vzh,Mergulhao:2023ukp,Euclid:2023shr,Calderon:2025xod}, future CMB polarisation observations will provide an exceptionally clean window on the early Universe, with unique potential to confirm or decisively exclude the presence of primordial features~\cite{Mortonson:2009qv,Finelli:2016cyd,Hazra:2017joc,Braglia:2022ftm}.

In this work, we perform a comprehensive search for primordial oscillatory features in \textit{Planck} data. We start by analysing the \textit{Planck} 2018 legacy release (PR3)~\cite{Planck_likelihood_2020,Efstathiou_2021}, and we also consider the \textit{Planck} PR4 (\texttt{NPIPE}) dataset~\cite{camspec_npipe_2022}, which is based on a new processing pipeline designed to produce calibrated frequency maps with reduced noise~\cite{Planck:2020olo}.
Since primordial features can manifest as high-frequency oscillations in multipole space, the standard binning procedures commonly adopted in cosmological parameter estimation may smooth out or dilute the corresponding signal. For this reason, an unbinned analysis is essential for assessing the stability of candidate high-frequency oscillations.
We compare the results obtained from these datasets for four different feature templates. Finally, looking ahead to the next generation of observations, we forecast the sensitivity of future experiments such as the Simons Observatory (SO)~\cite{SimonsObservatory:2018koc} and LiteBIRD~\cite{LiteBIRD:2022cnt}, emphasising the transformative role that high-precision polarisation data is expected to play in this field.

Ground-based CMB measurements, such as those from ACT~\cite{AtacamaCosmologyTelescope:2025blo} and SPT~\cite{SPT-3G:2025bzu}, extend the observed multipole range to significantly smaller angular scales. These measurements have a higher sensitivity and angular resolution than those from \textit{Planck} and therefore provide valuable additional information for inferring standard cosmological parameters. In the present case, however, we do not expect them to lead to a significant improvement in the feature constraints obtained from the unbinned \textit{Planck} likelihoods. This is because the relevant high-multipole likelihoods are based on band power spectra with a minimum bin width of $\Delta \ell = 50$, which increases towards higher multipoles. This smooths out percent-level oscillatory signatures and reduces sensitivity to extended oscillatory templates. This expectation is consistent with previous analyses combining \textit{Planck} with earlier SPT-3G data, where the addition of small-scale measurements resulted in tighter constraints for constant-amplitude oscillatory templates but did not lead to a significant qualitative improvement, although some improvement is naturally expected for features localised at small scales, see Ref.~\cite{Antony:2024vrx}. For these reasons, we focus in this work only to \textit{Planck} datasets and likelihoods. 

The structure of the paper is organised as follows. In~\cref{sec:theory}, we review the feature templates tested in this work, while~\cref{sec:data} describes the datasets and analysis pipeline. \Cref{sec:results} presents our main results from current observations. In~\cref{sec:forecasts}, we assess the sensitivity of future surveys, and we conclude in~\cref{sec:conclusions}. Details of the analysis and supplementary results are deferred to the appendices.

\section{Templates for primordial scale-dependent oscillatory features} \label{sec:theory}
We assume that the PPS can be written as 
    \begin{equation} \label{eqn:prim_spectrum_split}
    P_\zeta(k) = P_{\zeta,0}(k) \left[1 + \delta P^X_\zeta(k)\right] \,,
\end{equation}
where the featureless component of the scalar PPS is
\begin{align}
    P_{\zeta, 0}(k) &\equiv \frac{2 \pi^2}{k^3} \Delta_{\zeta, 0}^2(k) \notag\\
    &\equiv \frac{2 \pi^2}{k^3} A_\mathrm{s} \left(\frac{k}{k_*}\right)^{n_\mathrm{s}-1} \,,
\end{align}
with $\Delta_{\zeta, 0}^2(k)$ corresponding to the standard nearly scale-invariant PPS, where $A_\mathrm{s}$ and $n_\mathrm{s}$ denote the scalar amplitude and spectral index at the pivot scale $k_*$, which we fix to $k_* = 0.05 \, \mathrm{Mpc}^{-1}$. The quantity $\delta P_\zeta^X(k)$ encodes deviations from the power-law PPS, namely the primordial features.

\begin{figure}
    \centering
    \includegraphics[width=\linewidth]{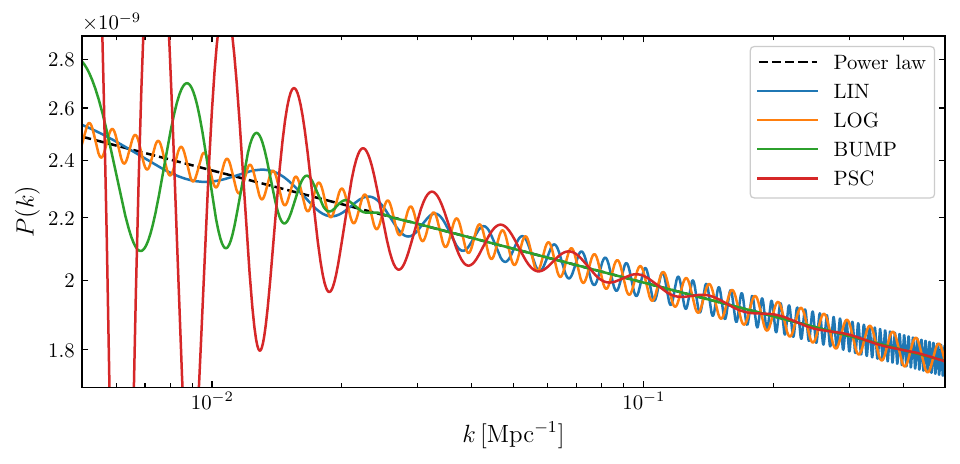}
    \caption{Primordial power spectrum of curvature perturbations for LIN, LOG, BUMP, and PSC models with best-fit parameters from~\cref{tab:bestfits}.}
    \label{fig:PPS}
\end{figure}

In the presence of primordial features with linearly and logarithmically spaced oscillations, we consider the following generic parametrisation entering~\cref{eqn:prim_spectrum_split}
\begin{equation} \label{eqn:Pk_template}
    \delta P_\zeta^\mathrm{X}(k) = A_\mathrm{X}(k) \sin \left(\omega_\mathrm{X} \Xi_\mathrm{X} + \phi_\mathrm{X}\right) \,,
\end{equation}
where $\Xi_\mathrm{X} = k/k_*$ for $X=\mathrm{lin}$ and $\Xi_\mathrm{X} = \ln(k/k_*)$ for $X=\mathrm{log}$. We refer to these as \textit{linear features} (LIN) and \textit{logarithmic features} (LOG), respectively. Here $A_\mathrm{X}(k)$ denotes the (possibly scale-dependent) amplitude or envelope of the primordial feature, $\omega_\mathrm{X}$ its dimensionless frequency, and $\phi_\mathrm{X}$ a phase. In the following, we describe the models considered in our analysis (the PPS for the templates considered is shown in~\cref{fig:PPS}).

\subsection{Sharp feature signal}
Linear oscillations arise in the PPS when a temporary departure from slow-roll occurs over a sufficiently short time interval; see Refs.~\cite{Starobinsky:1992ts,Adams:2001vc,Chen:2006xjb}. In this case, \textit{sharp} features produce an oscillatory signal in the PPS. 
In general, the amplitude is not constant but instead follows an envelope determined by the physical properties of the feature, such as its type, steepness, and duration; see, e.g., Ref.~\cite{Chluba:2015bqa} for a general review, while here we consider four representative templates. The envelope and phase of these oscillations depend on the inflationary potential and the background evolution, making them highly model-dependent. In multi-field scenarios, interactions with heavy fields can further modify both the amplitude and the phase, introducing additional complexity; see Refs.~\cite{Achucarro:2010da,Gao:2012uq,Chen:2014cwa}. In the following, we consider four representative scenarios with scale-independent amplitude and phase, which capture the essential physical behaviors while allowing direct comparison with previous analyses \cite{Akrami:2018odb,Braglia:2022ftm,Hamann:2021eyw}.

\subsection{Resonant feature signal}
Logarithmic oscillations arise when the slow-roll potential is subject to a periodic modulation, which induces an oscillatory correction to the background evolution; see Refs.~\cite{Chen:2008wn,McAllister:2008hb,Flauger:2009ab}. The modulation induces background oscillations with a fixed frequency in time. Each Fourier mode is resonantly enhanced when its physical frequency $k/a(t)$ matches this oscillation frequency.
As a result, each mode undergoes a transient \textit{resonant} phase while still inside the horizon, leaving a distinctive oscillatory imprint on the PPS. In this class of models, the modulation typically retains an approximately constant amplitude across scales.

\subsection{Bump/dip in the potential}
As a physically motivated template with linear oscillations, models containing a step, dip, or bump in the inflationary potential generate localised oscillations through a brief violation of slow-roll; see Refs.~\cite{Starobinsky:1992ts,Adams:2001vc}. Sharp turns in multi-field inflation~\cite{Achucarro:2010da} produce similar effects, with oscillations associated with changes in the inflationary trajectory. In addition, sudden variations in the speed of sound~\cite{Bean:2008na} give rise to oscillatory signatures with a scale-dependent envelope. These models differ in their oscillation phase, envelope, and damping behaviour.  Among the different possible features within this class, we adopt as a representative case the \textit{bump/dip template} (BUMP) computed in Ref.~\cite{Braglia:2022ftm} using the \textit{in-in} formalism~\cite{Chen:2010bka}\footnote{We note that the correction to the near-scale-invariant PPS considered in this paper is computed at tree-level in the in-in formalism. Although loop corrections alter this scale dependence, they are negligible for the parameters explored in this paper; see Ref.~\cite{Braglia:2026fle}.}
\begin{equation} \label{eqn:BUMP_template}
    \delta P_\zeta^\mathrm{bump}(k) = A_\mathrm{bump} \sqrt{\frac{e}{2\alpha}} \left[2\cos(x_0) + \left(x_0 - \frac{2}{x_0}\right)\sin(x_0)\right] e^{-x_0^2/(4\alpha)} \,,
\end{equation}
where $x_0 \equiv k\,\omega_\mathrm{bump}/k_*$, so as to use the same notation as in \cref{eqn:Pk_template} for linear oscillations, and $\alpha$ is inversely proportional to the duration of the sharp feature, with $\Delta N = 36/\alpha^2$.

\subsection{Primordial standard clock signal}
Classical or quantum oscillations of massive fields can act as standard clocks, directly probing the time dependence of the scale factor in the primordial Universe; see Refs.~\cite{Chen:2011zf,Chen:2014cwa,Chen:2015lza}. In this work, we focus on 
a particular classical realization, in which a heavy field is excited by a sharp feature during inflation.

Following Ref.~\cite{Chen:2011zf}, we parametrise the scale factor in different primordial-Universe scenarios as $a(t) \sim t^p$, such that the value of $p$ distinguishes different backgrounds: $|p|>1$ corresponds to inflation, $0<p\sim \mathcal{O}(1)<1$ to fast contraction, $0<p\ll 1$ to slow contraction, and $-1\ll p<0$ to slow expansion.

We focus on the leading contribution to the primordial standard clock signal (PSC) generated in this sharp-excitation regime. 
For inflationary models with $p\gg 1$, the parameter $p$ is degenerate with the phase $\phi_\mathrm{psc}$, so in this limit 
\begin{align}
  \delta P_\zeta^\mathrm{psc}(k) = \begin{cases}
    0 \,, & \text{if $k <  k_r$} \,,  \\
    A_\mathrm{psc} \left( \frac{k}{k_r}  \right)^{-\frac{3}{2}}
    \sin \left[ \omega_\mathrm{psc} \ln \left(\frac{k}{k_r}\right) + \phi_\mathrm{psc} \right] \,,  & \text{otherwise} \,.
  \end{cases}
  \label{eqn:PSC_template}
\end{align}

\section{Cosmological datasets and analysis pipeline} \label{sec:data}
This section describes the datasets, numerical tools, and analysis settings used for parameter inference and statistical model comparison.

Our analysis focuses on CMB measurements from the \textit{Planck} PR3 release and its updated reanalysis based on PR4 \texttt{NPIPE} maps. For all CMB analyses, we adopt the same low-$\ell$ and lensing likelihood combination: \texttt{Commander} for large-scale temperature anisotropies, \texttt{SRoll2} for low-$\ell$ $E$-mode polarisation~\cite{Delouis_2019,Pagano:2019tci}, and the \textit{Planck} PR4 lensing likelihood~\cite{Carron_2022}.
For the high-$\ell$ CMB data, we consider three different likelihoods:
\begin{itemize}
    \item \textit{Plik-PR3}: the unbinned version of the \textit{Planck} PR3 (2018) legacy likelihood~\cite{Planck_likelihood_2020}. Unlike the standard binned version, which averages the power spectra over multipole bins (typically $\Delta \ell \sim 30$), this likelihood retains the full multipole-by-multipole resolution of the temperature and polarisation spectra. This is particularly important for our analysis, since binning can smooth or partially erase the sharp oscillatory patterns predicted by high-frequency feature models. The likelihood uses cross-spectra between half-mission maps at 100, 143, and 217\,GHz.
    \item \textit{CamSpec-PR3}: the \texttt{CamSpec} likelihood applied to the \textit{Planck} PR3 (2018) legacy maps~\cite{Planck_likelihood_2020,Efstathiou_2021}. Relative to \texttt{Plik}, this likelihood adopts different masks, a different treatment of Galactic dust foregrounds based on 545\,GHz maps, and a different strategy for combining cross-spectra. Including this likelihood allows us to test the robustness of our results with respect to alternative high-$\ell$ likelihood constructions applied to the same PR3 maps.
    \item \textit{CamSpec-PR4}: the \texttt{CamSpec} high-$\ell$ likelihood~\cite{Efstathiou_2021,camspec_npipe_2022} based on the \textit{Planck} PR4 \texttt{NPIPE} maps~\cite{Planck:2020olo}. The \texttt{NPIPE} processing pipeline reanalyses the time-ordered data, yielding maps with lower noise levels and better-characterised systematics than the PR3 release. Within our current dataset selection, this constitutes the most constraining \textit{Planck}-based configuration.
\end{itemize}

Given the highly multimodal nature of the parameter space associated with high-frequency primordial features, the parameter inference is performed using the nested-sampling algorithm implemented in \texttt{PolyChord}~\cite{Polychord1_Handley_2015,Polychord2_Handley_2015}, with \texttt{Cobaya}~\cite{Cobaya_Torrado_2021} used as the framework for model specification and likelihood evaluation. 
Posterior distributions, parameter means, and confidence levels (CLs) are computed and visualised using \texttt{GetDist}~\cite{Lewis:2019xzd}.
Unlike standard Markov Chain Monte Carlo (MCMC) techniques, which may struggle to converge in multimodal landscapes, nested sampling is specifically designed to efficiently explore isolated modes and provides a robust estimation of the Bayesian evidence, which is crucial for our model selection.
In all runs, the number of live points is set to 1000 in order to ensure a robust exploration of the posterior distribution. Theoretical predictions for the CMB anisotropies are computed with a modified version of \texttt{CAMB}~\cite{CAMB_Lewis_2000,Howlett_2012}. The primordial helium abundance is consistently derived using the public results generated with \texttt{PArthENoPE}~\cite{Gariazzo:2021iiu}, accounting for the variation of the baryon density parameter $\omega_{\rm b}$. For the analysis of the best-fits we use \texttt{py-BOBYQA}~\cite{10.1145/3338517} implementation in \texttt{Cobaya} as minimizer and we perform a maximum a posteriori (MAP) analysis varying all cosmological parameters, nuisance parameters included.

The priors adopted for the standard cosmological parameters and for the feature parameters are reported in~\cref{tab:priors_cmb}. In all cases, we simultaneously vary the nuisance parameters associated with residual foregrounds and instrumental systematics, as defined for each likelihood.

\begin{table}[t]    
    \small
    \centering
    \renewcommand{\arraystretch}{1.5} 
    \begin{tabular}{l c}
    \hline\hline
    \textbf{Parameter} & \textbf{Prior} \\
    \hline
    $\omega_\mathrm{c}$ & [0.095,\,0.145] \\
    $\omega_\mathrm{b}$ & [0.019,\,0.025] \\
    $\tau_\text{reio}$ & [0.01,\,0.3] \\
    $100\,\theta_\mathrm{MC}$ & [1.03,\,1.05] \\
    $\ln\left(10^{10}A_\mathrm{s}\right)$ & [1.61,\,3.91] \\
    $n_\mathrm{s}$ & [0.8,\,1.2] \\
    \hline\hline
    \end{tabular}
    \hspace{0.5cm}
    \begin{tabular}{l c c c c}
    \hline\hline
    \textbf{Parameter} & \textbf{LIN} & \textbf{LOG} & \textbf{BUMP} & \textbf{PSC} \\
    \hline
    $A_X$ & $[0,\,0.5]$ & $[0,\,0.5]$ & $[-0.5,\,0.5]$ & $[0,\,0.5]$ \\
    $\log_{10}\omega_X$ & $[0,\,2.1]$ & $[0,\,2.1]$ & $[0,\,2.1]$ & $[0,\,2.1]$ \\
    $\phi_X/2\pi$ & $[0,\,1]$ & $[0,\,1]$ & -- & $[0,\,1]$ \\
    $\log_{10}\alpha$ & -- & -- & $[1.5,\,4]$ & -- \\
    $\log_{10} k_r$ & -- & -- & -- & $[-2.3,\,-0.8]$ \\
    \hline\hline
    \end{tabular}
    \caption{Uniform priors adopted for the standard cosmological parameters (left) and feature parameters (right).}
    \label{tab:priors_cmb}
\end{table}

\section{Current CMB constraints on primordial feature templates} \label{sec:results}
In this section, we present the results obtained for the four feature templates described in~\cref{sec:theory} using the three CMB likelihood combinations described in~\cref{sec:data}.
In~\cref{fig:triangle_features}, we show the marginalised posterior distributions for the extra parameters of the different feature templates. For a given likelihood combination, the standard cosmological parameters remain consistent across the various feature parametrisations; the corresponding constraints are reported in~\cref{app:cosmo_params}.

\subsection{Global constraints on the feature amplitude}
\begin{figure}
    \includegraphics[scale=0.5]{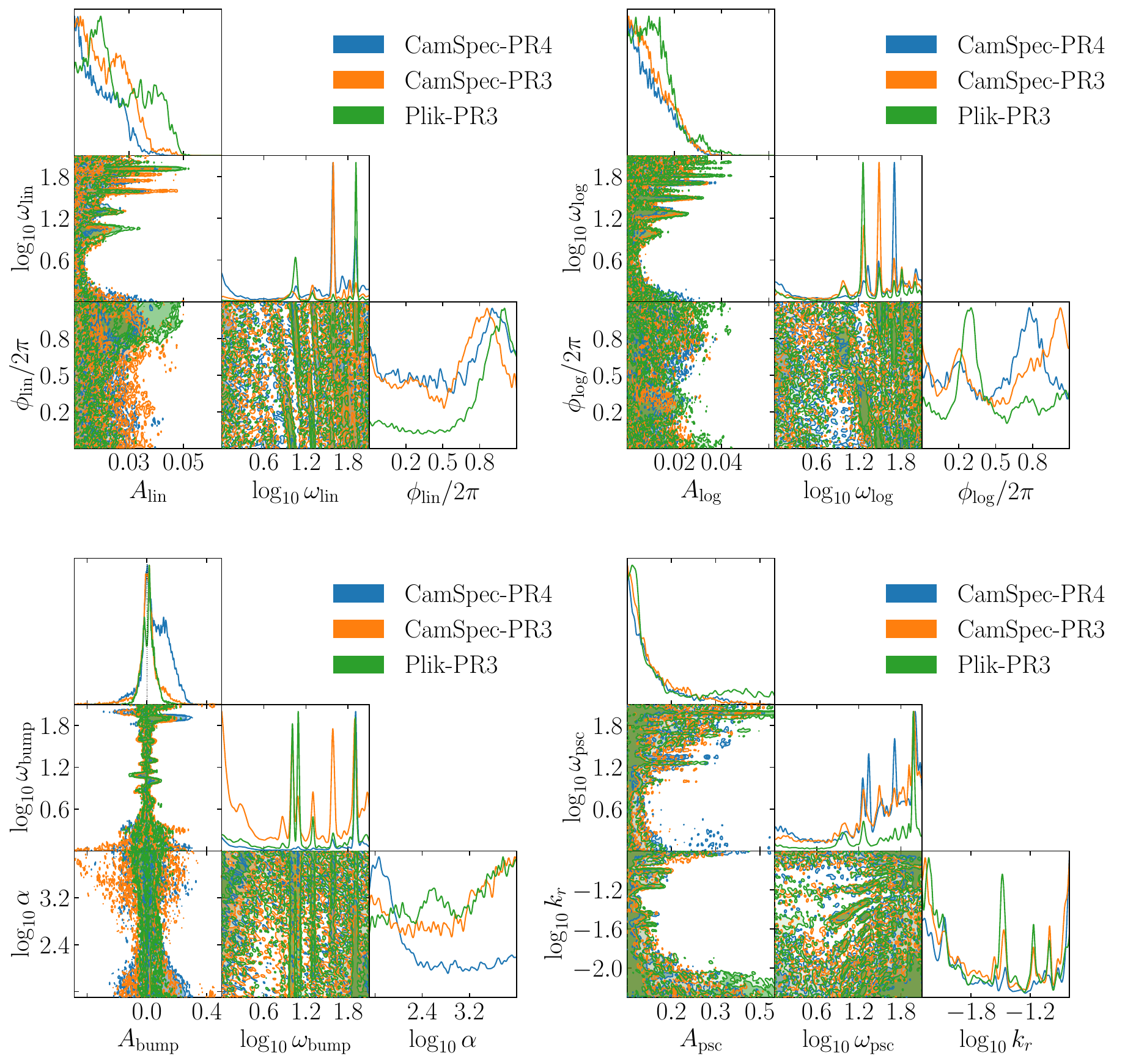}
    \caption{Marginalised joint 68\% and 95\% CL posterior distribution for feature parameters obtained using the LIN, LOG, BUMP, and PSC templates for the different datasets (\textit{Plik-PR3} in green, \textit{CamSpec-PR3} in orange, and \textit{CamSpec-PR4} in blue). We do not show the Marginalised posterior for $\phi_\mathrm{psc}$, as it is completely flat.}
    \label{fig:triangle_features}
\end{figure}

From the posterior distributions shown in~\cref{fig:triangle_features}, we derive the amplitude constraints reported in \cref{tab:amplitudes}. For the LOG template, the upper bounds on the feature amplitude are broadly consistent across the three likelihood combinations. For the LIN template, the constraints are likewise stable overall, although \textit{Plik-PR3} yields a slightly weaker upper bound than both \textit{CamSpec-PR3} and \textit{CamSpec-PR4}. The BUMP template is also consistently constrained, with only a mild dependence on the adopted likelihood combination. 

The broader allowed ranges for the amplitudes of the BUMP and PSC templates are partly due to the different localisation of these oscillatory templates relative to the LIN and LOG cases. 
This effect is more pronounced for the PSC template because of the additional parameter $k_r$, which controls the characteristic scale at which the signal is activated. 
The prior range adopted for $k_r$ is motivated by the region of the PPS that is constrained by CMB data at the level of a few per cent~\cite{Akrami:2018odb,Handley:2019fll,Raffaelli:2025kew}. 
In the inflationary PSC template, however, the oscillatory signal is not uniformly distributed, since for $k \geq k_r$ it is modulated by a damping envelope scaling as $(k/k_r)^{-3/2}$. When $k_r$ lies near the edges of the prior range, a substantial fraction of the oscillatory contribution is either pushed outside the best-constrained CMB window or strongly suppressed within it.
A similar consideration applies to the BUMP template. In this case, the characteristic location of the feature scales as $\propto \sqrt{\alpha}$. The adopted prior on $\alpha$ maps into a  narrower interval in effective $k$-space than the prior adopted for $k_r$. The two prior choices are thus only approximately comparable.

Overall, our amplitude constraints for \textit{Plik-PR3} are consistent with those reported in Refs.~\cite{Akrami:2018odb,Hamann:2021eyw,Braglia:2022ftm,Calderon:2025xod}, whenever a direct comparison is possible.

\begin{table}[t]
    \small
    \centering
    \renewcommand{\arraystretch}{1.4} 
    \begin{tabular}{ccccc}
    \hline\hline
    \textbf{Dataset}    & \boldmath $A_\mathrm{lin}$ & \boldmath $A_\mathrm{log}$ & \boldmath $A_\mathrm{bump}$ & \boldmath $A_\mathrm{psc}$ \\
    \hline
    \textit{Plik-PR3}    & $<0.042$ & $<0.028$ & $_{-0.09}^{+0.13}$ &  $<0.44^*$ \\
    \textit{CamSpec-PR3} & $<0.031$ & $<0.025$ & $_{-0.18}^{+0.17}$ & $<0.26$ \\
    \textit{CamSpec-PR4} & $<0.027$ & $<0.025$ & $_{-0.11}^{+0.25}$ & $<0.29$ \\
    \hline\hline
    \end{tabular}
    \caption{Marginalised constraints on the amplitudes of the oscillatory feature templates for the different datasets. Bounds are quoted at 95\% CL, unless otherwise stated. *The 95\% CL bound for the amplitude is obtained integrating the one-dimensional posterior distribution from the left-hand side over the prior range $[0,0.5]$.}
    \label{tab:amplitudes}
\end{table}

\subsection{Preferred frequencies and best-fit improvements}
Global amplitude constraints provide a useful summary of how strongly each template is constrained once the full frequency range has been marginalised over. However, they do not capture the presence of specific frequencies at which a given oscillatory template may provide a better fit to the CMB data than featureless PPS. For this reason, we complement the posterior analysis with a dedicated study of the preferred frequencies, aimed at identifying the parameter values that yield the largest improvement in fit for each template and likelihood combination.
To this end, we analyse the quantity $\Delta\chi^2 \equiv \chi^2_X - \chi^2_{\Lambda\mathrm{CDM}}$, where $\chi^2_{\Lambda\mathrm{CDM}}$ denotes the best-fit value obtained under featureless PPS and $\chi^2_X$ is the corresponding best-fit value for the template with oscillatory features. 

\begin{figure}
    \centering
    \includegraphics[width=\linewidth]{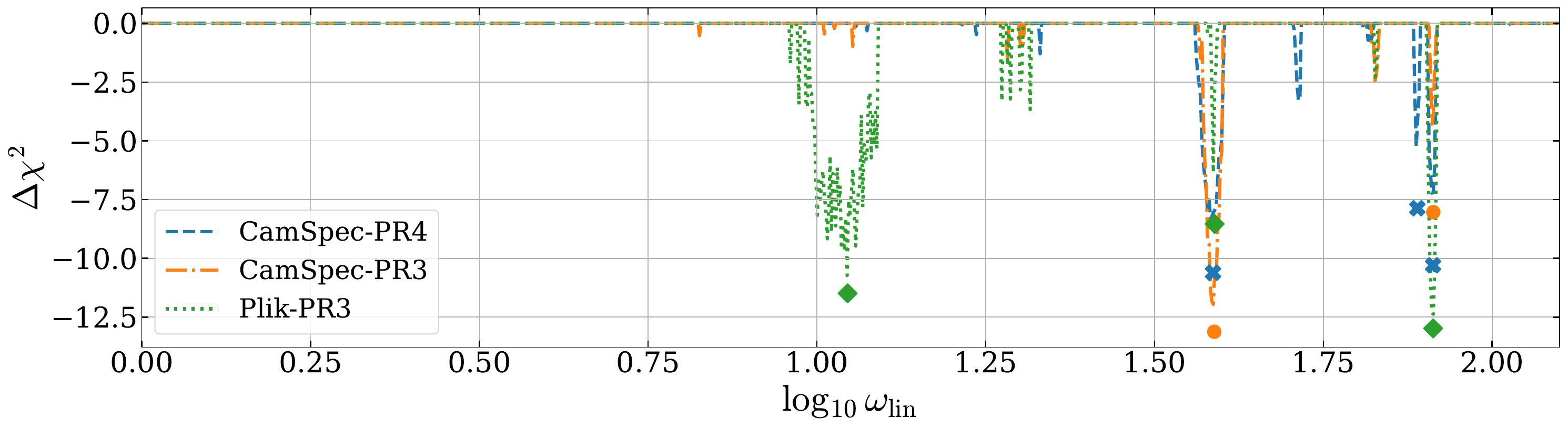}
    \includegraphics[width=\linewidth]{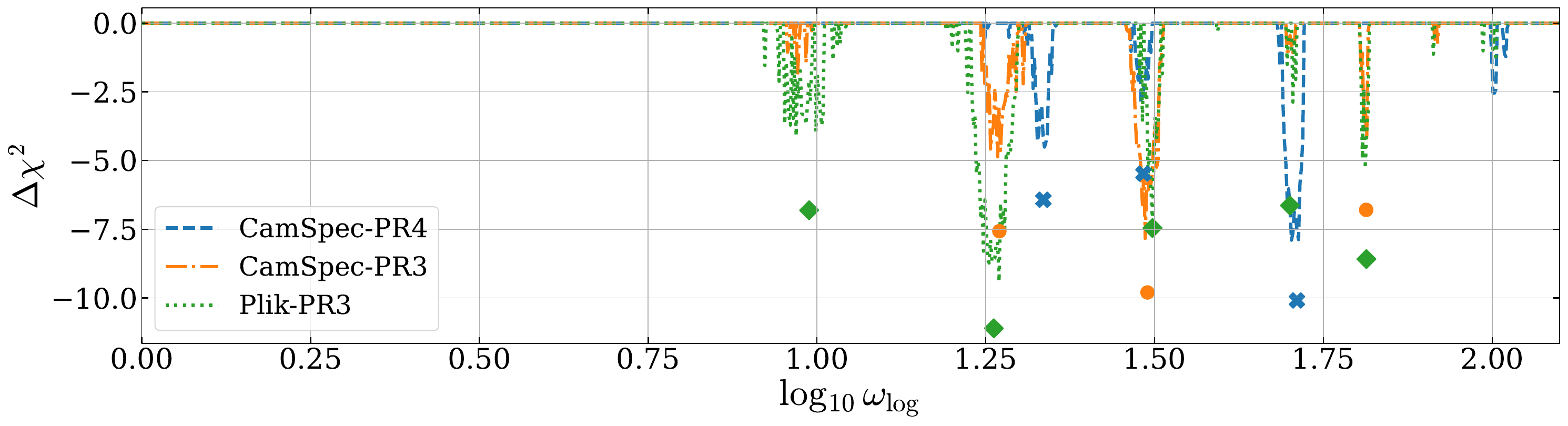}
    \includegraphics[width=\linewidth]{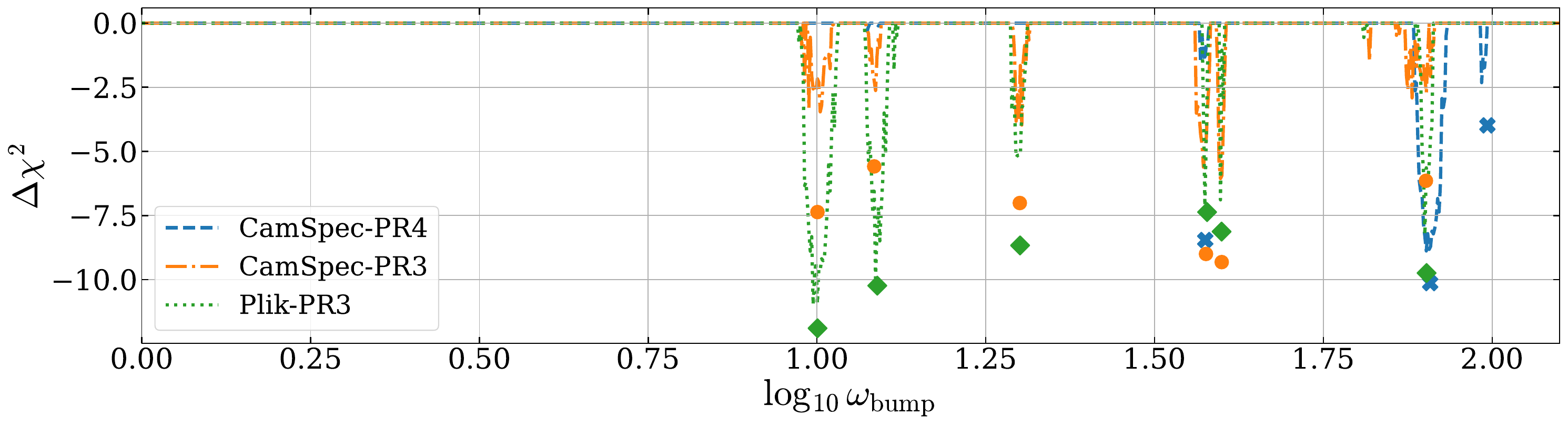}
    \includegraphics[width=\linewidth]{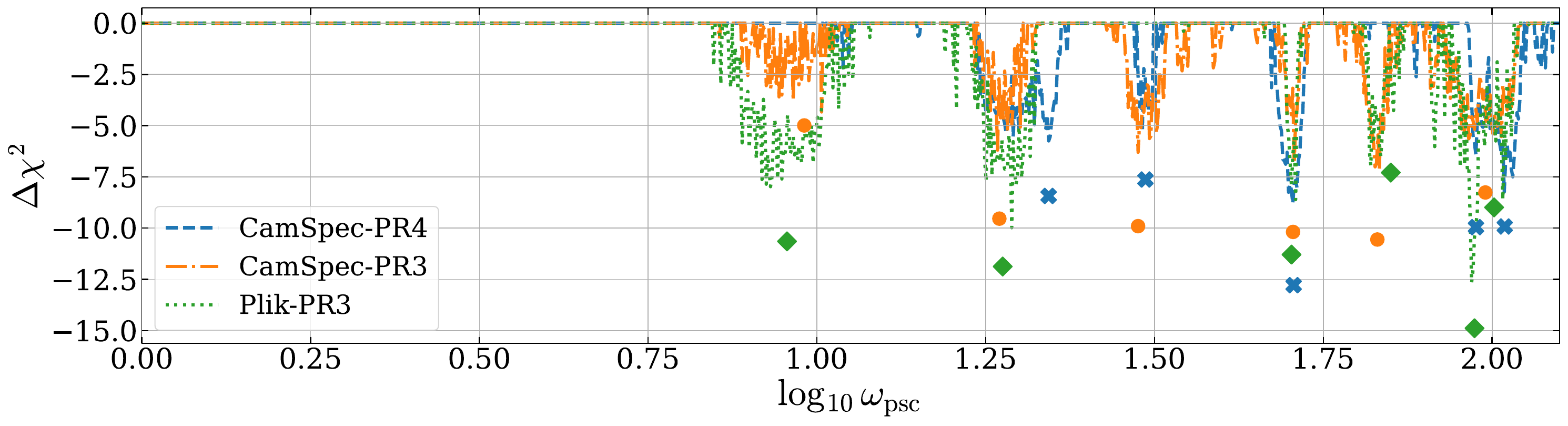}
    \caption{Sample-based reconstruction of $\Delta\chi^2$ as a function of the feature frequency for the different likelihood combinations (\textit{Plik-PR3} in green, \textit{CamSpec-PR3} in orange, and \textit{CamSpec-PR4} in blue). Here $\Delta\chi^2 \equiv \chi^2_X - \chi^2_{\Lambda\mathrm{CDM}}$, so negative values correspond to an improved fit relative to featureless $\Lambda$CDM. The curves are obtained directly from the likelihood samples, with all remaining parameters varied consistently, while the markers indicate the local best-fit values found with the dedicated minimiser around the dominant peaks.
    The different panels correspond, from top to bottom, to the LIN, the LOG, the BUMP, and the PSC templates, respectively.}
    \label{fig:profile}
\end{figure}

We first examine the sample-based variation of $\Delta\chi^2$ as a function of frequency  in order to identify the main candidate frequencies, see~\cref{fig:profile}. We then refine these candidate solutions with a minimiser, which provides the corresponding local best-fit parameter values and best-fit $\Delta\chi^2$. The resulting best-fit frequencies for the different templates and likelihood combinations are reported in \cref{tab:bestfits}, while the remaining local best-fits are listed in~\cref{app:bestfits}.
In~\cref{fig:profile}, we show the improvement in fit with respect to the baseline featureless model, as estimated from the samples of our analysis, as a function of the feature frequency, with all remaining parameters varied consistently. We additionally overplot the precise values of $\Delta\chi^2$ obtained with the minimiser around the main peaks identified in the posterior distributions.

For \textit{Plik-PR3}, our results partially confirm the analysis of Ref.~\cite{Akrami:2018odb}. 
For the LIN template, we do find a peak close to the frequency reported in Ref.~\cite{Akrami:2018odb}, namely $\log_{10}\omega_\mathrm{lin} = 1.05$, but this is not the global best-fit solution in our analysis. Instead, in agreement with Ref.~\cite{Hamann:2021eyw}, the preferred solution is shifted towards higher frequencies, with a best fit at $\log_{10}\omega_\mathrm{lin} = 1.91$.
For the LOG template, we recover the same global best-fit frequency as in Ref.~\cite{Akrami:2018odb}, $\log_{10}\omega_\mathrm{log} = 1.26$, together with a comparable improvement in fit. This preferred frequency was also identified in Refs.~\cite{Braglia:2022ftm,Hamann:2021eyw}.
The best-fit frequencies identified for the BUMP template also appear to be broadly consistent with the analysis of Ref.~\cite{Braglia:2022ftm}, although explicit numerical values are not quoted there and the analysis was performed numerically without relying on an analytical templates. 
For the PSC template, our best-fit frequency agrees with that found in Ref.~\cite{Braglia:2022ftm}. More generally, we find that the preferred frequencies of the PSC template are qualitatively aligned with those obtained for the LOG template, while the frequencies favoured by the BUMP template are broadly compatible with those found for the LIN template.

The best-fit frequency shifts in almost all cases when using \textit{Plik-PR3}, \textit{CamSpec-PR3} or \textit{CamSpec-PR4}. Nevertheless, the preferred frequencies remain in the range $10 \lesssim \omega \lesssim 100$, yielding improvements of $\Delta\chi^2 \sim -10$ to $-15$. 
The primordial feature best-fit parameters associated with the main posterior peaks, together with their corresponding improvements in $\Delta \chi^2$, are reported in~\cref{tab:osc_params_1,tab:osc_params_2,tab:osc_params_3,tab:osc_params_4}.

\begin{table}[t]
    \small
    \centering
    \renewcommand{\arraystretch}{1.4} 
    \setlength{\tabcolsep}{5pt}
    \begin{tabular}{cccccccccc}
    \hline\hline
    \textbf{Model} & \textbf{Dataset} & \boldmath $A_X$ & \boldmath $\log_{10}\omega_X$ & \boldmath $\phi_X/2\pi$ & \boldmath $\log_{10}\alpha$ & \boldmath $\log_{10}k_r$ & \boldmath $\Delta\chi^2$ & \boldmath $\ln B$ & $\bm{S}$ \\
    \hline
    \multirow[c]{3}{*}{LIN}
    & \emph{Plik-PR3}      & $0.035$ & $1.91$ & $0.88$ & -- & -- & $-12.7$ & $-2.4$ & $2.0$ \\[-4pt]
    & \emph{CamSpec-PR3}   & $0.027$ & $1.59$ & $0.76$ & -- & -- & $-13.1$ & $-2.9$ & $2.2$ \\[-4pt]
    & \emph{CamSpec-PR4}   & $0.023$ & $1.58$ & $0.84$ & -- & -- & $-10.6$ & $-3.7$ & $1.7$ \\
    \hline
    \multirow[c]{3}{*}{LOG}
    & \emph{Plik-PR3}      & $0.015$ & $1.26$ & $0.31$ & -- & -- & $-11.1$ & $-3.1$ & $1.5$ \\ [-4pt]
    & \emph{CamSpec-PR3}   & $0.019$ & $1.49$ & $0.94$ & -- & -- & $-9.8$ & $-2.9$ & $1.2$ \\ [-4pt]
    & \emph{CamSpec-PR4}   & $0.024$ & $1.71$ & $0.72$ & -- & -- & $-10.1$ & $-3.6$ & $1.3$ \\
    \hline
    \multirow[c]{3}{*}{BUMP}
    & \emph{Plik-PR3}      & $0.17$ & $1.00$ & -- & $1.72$ & -- & $-11.9$ & $-2.1$ & $1.9$ \\[-4pt]
    & \emph{CamSpec-PR3}   & $-0.052$ & $1.60$ & -- & $4.00$ & -- & $-9.3$ & $-2.3$ & $1.1$ \\[-4pt]
    & \emph{CamSpec-PR4}   & $0.14$ & $1.90$ & -- & $1.78$ & -- & $-10.1$ & $-1.7$ & $1.4$ \\
    \hline
    \multirow[c]{3}{*}{PSC}
    & \emph{Plik-PR3}      & $0.50$ & $1.97$ & $0.24$ & -- & $-2.23$ & $-14.9$ & $-2.0$ & $2.6$ \\[-4pt]
    & \emph{CamSpec-PR3}   & $0.076$ & $1.83$ & $0.17$ & -- & $-1.17$ & $-10.6$ & $-0.6$ & N/A \\[-4pt]
    & \emph{CamSpec-PR4}   & $0.070$ & $1.70$ & $0.85$ & -- & $-1.55$ & $-12.8$ & $-0.5$ & $0.7$ \\
    \hline\hline
    \end{tabular}
    \caption{Best-fit values of the template parameters. We also report $\Delta \chi^2 \equiv \chi^2_X - \chi^2_{\Lambda \mathrm{CDM}}$, $\ln B \equiv \ln\!\left(\mathcal{Z}_X / \mathcal{Z}_{\Lambda \mathrm{CDM}}\right)$, and the look-elsewhere-corrected significance $S$ reported in units of $\sigma$ according to~\cref{eq:lee_significance}. Negative values of $\Delta \chi^2$ correspond to a better fit of the featureless PPS as well as positive values of $\ln B$ do.}
    \label{tab:bestfits}
\end{table}

\begin{figure}
    \centering
    \includegraphics[width=\linewidth]{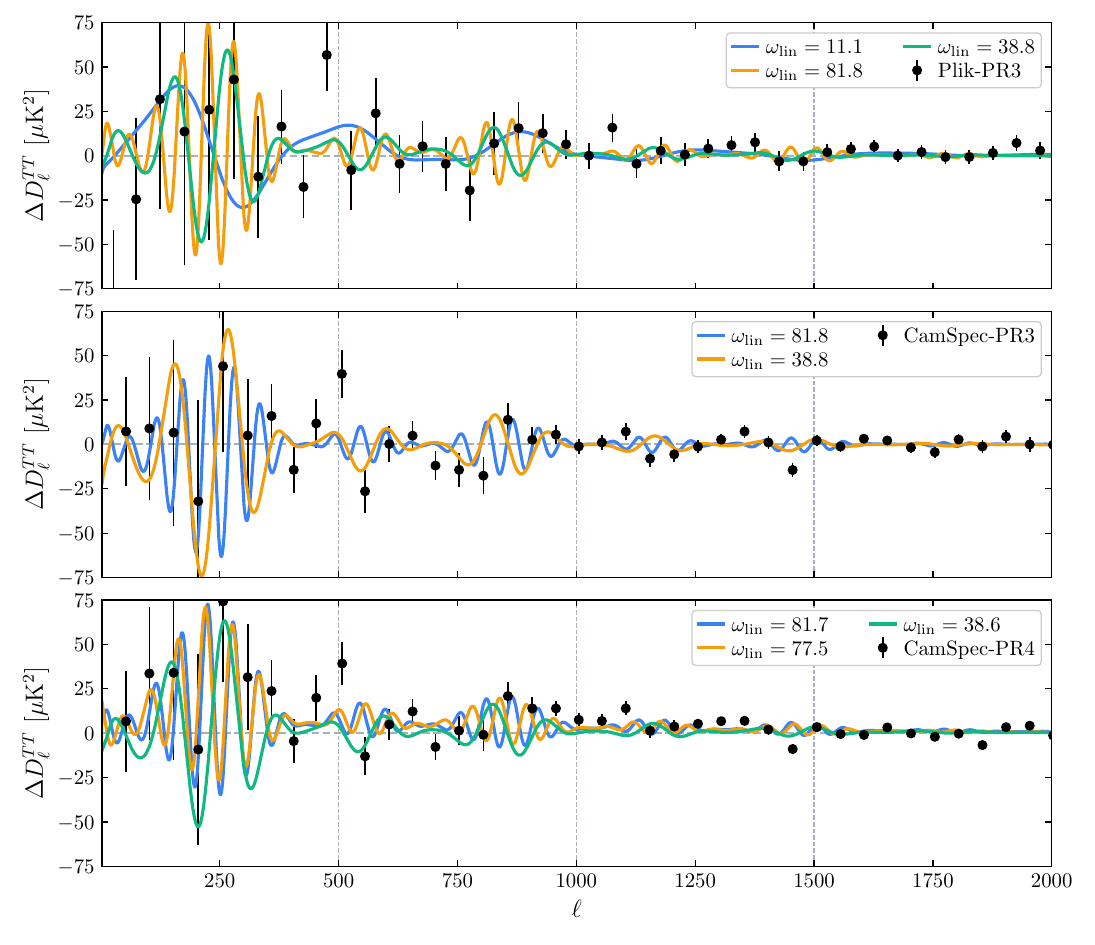}
    \caption{Residuals of the $TT$ angular power spectrum with respect to the best-fit featureless PPS reference model for the LIN template. The black points with error bars show the residuals of the measured bandpowers relative to the corresponding binned $\Lambda$CDM prediction, while the coloured curves show the residuals of the best-fit LIN solutions with respect to the same reference. From top to bottom, the panels correspond to \emph{Plik-PR3}, \emph{CamSpec-PR3}, and \emph{CamSpec-PR4}. The labels in the legend indicate the frequencies of the best-fit solutions.}
    \label{fig:res_TT_LIN}
\end{figure}

\begin{figure}
    \centering
    \includegraphics[width=\linewidth]{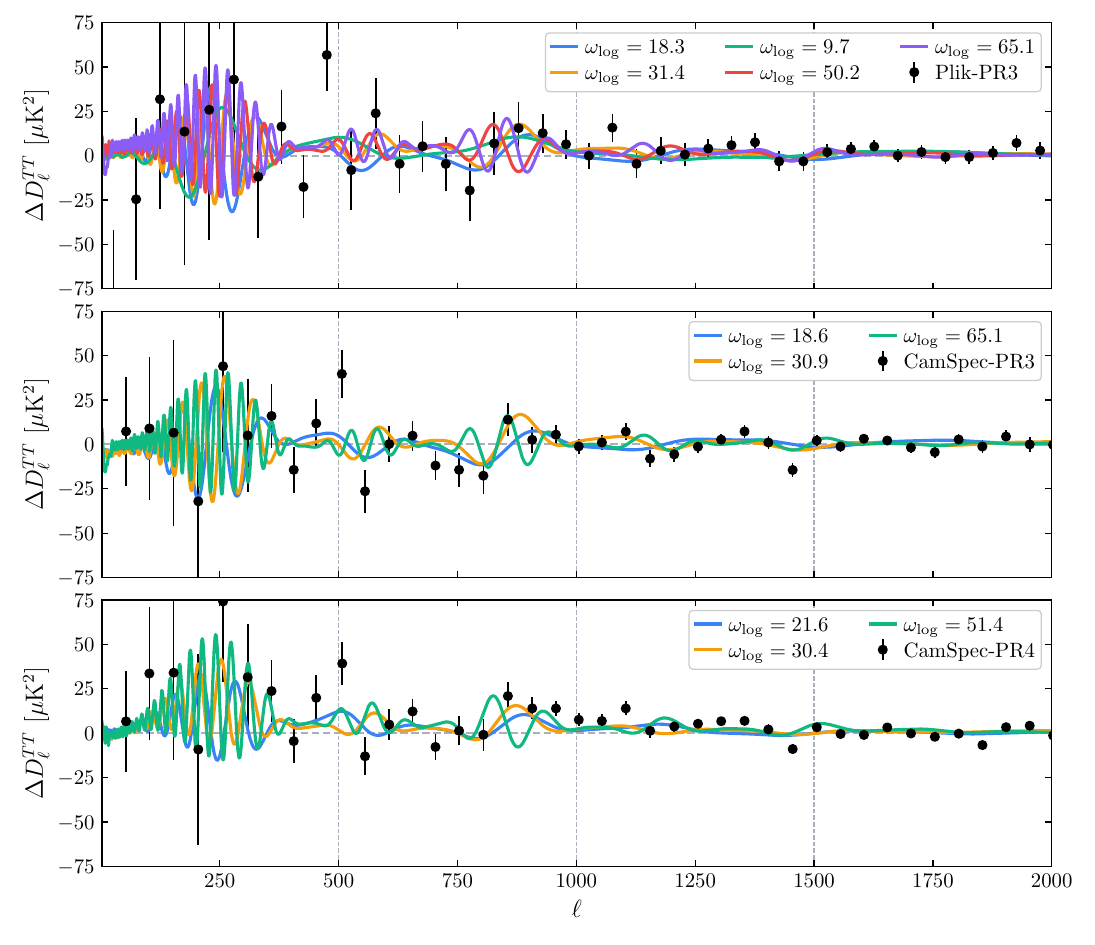}
    \caption{As in \cref{fig:res_TT_LIN}, but for the LOG template.}
    \label{fig:res_TT_LOG}
\end{figure}

\begin{figure}
    \centering
    \includegraphics[width=\linewidth]{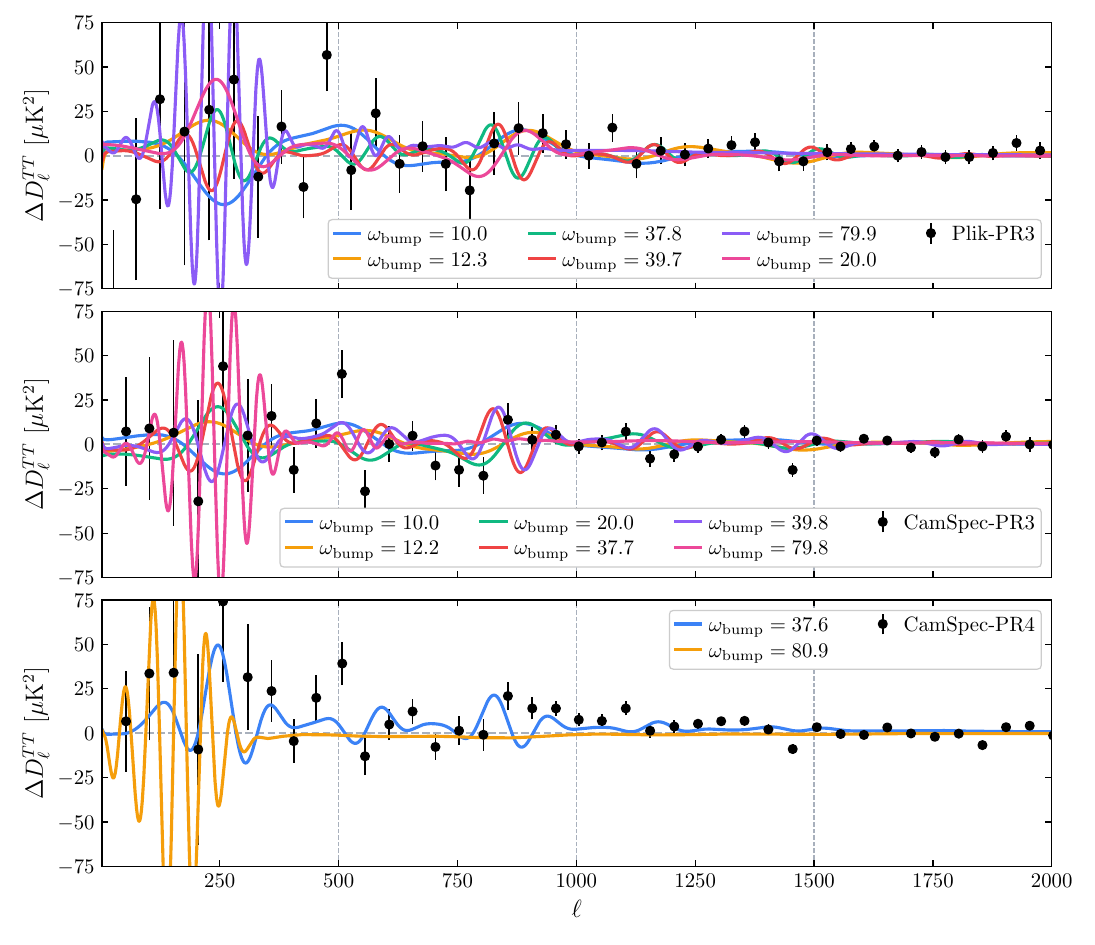}
    \caption{As in \cref{fig:res_TT_LIN}, but for the BUMP template.}
    \label{fig:res_TT_BUMP}
\end{figure}

\begin{figure}
    \centering
    \includegraphics[width=\linewidth]{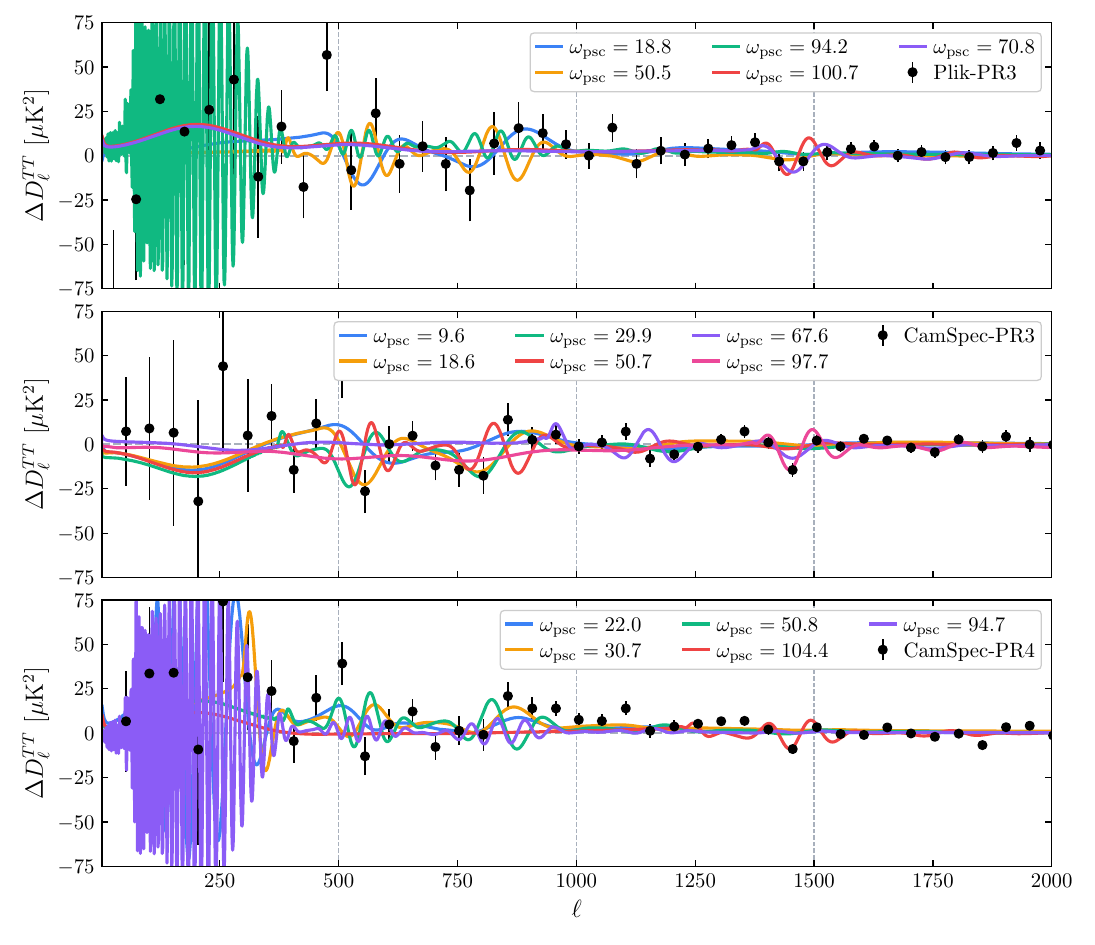}
    \caption{As in \cref{fig:res_TT_LIN}, but for the PSC template.}
    \label{fig:res_TT_CLOCK}
\end{figure}

To assess whether these improvements in fit are driven by specific structures in the data, we also inspect the residuals of the best-fit feature templates relative to the baseline featureless PPS case, following the approach of Ref.~\cite{Hamann:2021eyw}. In \cref{fig:res_TT_LIN,fig:res_TT_LOG,fig:res_TT_BUMP,fig:res_TT_CLOCK}, we show the residuals for the $TT$ angular power spectra, which dominate the analysis, while the corresponding $TE$ and $EE$ residuals are collected in \cref{app:residuals}. The binned data, with $\Delta \ell = 50$, are shown only as a visual guide to highlight the data points fitted by the different templates; all results and quoted statistical quantities are instead derived from the unbinned datasets.

For the LIN template, the improvement in fit is distributed over broad multipole intervals. The main contribution comes from $\ell \sim 1000$--$1500$, with an additional contribution at $\ell \sim 400$--$800$ for the lower-frequency solutions in \emph{Plik-PR3} and \emph{CamSpec-PR3}. Two frequency families, $\omega_\mathrm{lin} \sim 39$ and $\omega_\mathrm{lin} \sim 82$, are recovered across the three likelihoods. The global best-fit solution is found at $\omega_\mathrm{lin} \sim 81$ for \emph{Plik-PR3} and at $\omega_\mathrm{lin} \sim 38$ for both \emph{CamSpec} likelihoods. By contrast, the low-frequency solution is found only in \emph{Plik-PR3} likelihood.
For the LOG template, the improvement in fit gets two main contributions around $\ell \sim 900$--$1100$ and $\ell \sim 1300$--$1500$. Two low-frequency families, centred around $\omega_\mathrm{log} \sim 20$ and $\omega_\mathrm{log} \sim 30$, are recovered across the different likelihoods, while the high-frequency family, $\omega_\mathrm{log} \sim 50$--$65$, is less stable across datasets. The global best-fit frequency shifts from $\omega_\mathrm{log} \sim 18$ in Plik-PR3 to $\omega_\mathrm{log} \sim 32$ in \emph{CamSpec-PR3} and to $\omega_\mathrm{log} \sim 50$ in \emph{CamSpec-PR4}. 
In \emph{Plik-PR3}, two main contributions are present, around $\ell \sim 400$ and $\ell \sim 900$, with the latter mainly associated with the lower-frequency solutions, in particular $\omega_\mathrm{log} \sim 18$.
In \emph{CamSpec-PR3}, the dominant contribution is concentrated around $\ell \sim 800$. 
In \emph{CamSpec-PR4}, the gain is accumulated more gradually over $\ell \sim 800$--$1500$. 

For the BUMP template, the improvement in fit is more localised in multipole space than for the LIN and LOG templates. The main contribution arises between $\ell \sim 900$--$1500$, with an additional contribution at $\ell \sim 450$--$900$ for the low-frequency solutions. Two frequency families can be identified. The first is a low-frequency family, $\omega_\mathrm{bump} \sim 10$--$20$, which is supported mainly by the PR3 likelihoods. The second is centred around $\omega_\mathrm{bump} \sim 38$--$40$, which is recovered in all datasets. However, the global best-fit frequency shifts significantly across likelihoods, from $\omega_\mathrm{bump} \sim 10$ in \emph{Plik-PR3} to $\omega_\mathrm{bump} \sim 38$ in \emph{CamSpec-PR3}, and to $\omega_\mathrm{bump} \sim 81$ in \emph{CamSpec-PR4}. High-frequency solutions around $\omega_\mathrm{bump} \sim 80$--$100$ are therefore not reproduced consistently across the different likelihoods.
For the PSC template, the improvement in fit is distributed over broad but preferential multipole intervals, with the main contribution arising between $\ell \sim 900$--$1500$, and an additional contribution at $\ell \sim 500$--$900$. A low-frequency family around $\omega_\mathrm{psc} \sim 20$--$30$ is recovered in more than one likelihood, while a second, weaker family is present around $\omega_\mathrm{psc} \sim 50$--$70$. However, the global best-fit frequency is not stable across datasets, shifting from $\omega_\mathrm{psc} \sim 93$ in \emph{Plik-PR3} to $\omega_\mathrm{psc} \sim 68$ in \emph{CamSpec-PR3}, and to $\omega_\mathrm{psc} \sim 50$ in \emph{CamSpec-PR4}. Higher-frequency solutions around $\omega_\mathrm{psc} \sim 95$--$105$ are more dataset-dependent. In particular, the strongest \emph{Plik-PR3} solution receives a substantial contribution already at low multipoles and is not reproduced with comparable significance by the \emph{CamSpec} likelihoods.

In general, for \emph{CamSpec-PR4} the improvement in the fit is accumulated smoothly over the multipole range $\ell \sim 900$--$1500$, while the features present at $\ell \sim 500$--$800$ are better described by a featureless PPS. This is also reflected in the absence of local best-fit solutions with $\omega_X \lesssim 1.35$ when the \emph{CamSpec-PR4} likelihood is used.

\subsection{Bayes factor and look-elsewhere effect}
In addition to the values of the difference in chi-squared, we also report the global Bayes factor in~\cref{tab:bestfits}. In particular, we show $\ln B$ computed as $\ln B = \ln\mathcal{Z}_X - \ln\mathcal{Z}_{\Lambda\text{CDM}}$ where $\mathcal{Z}_{X}$ represents the evidence for a feature template given a dataset. The Bayes factor analysis favours the $\Lambda$CDM model, which represents the null hypothesis here, over oscillatory features in all cases, with $\ln B<-2$ in most cases. According to the revised Jeffreys' scale \cite{Trotta:2008qt}, a value of $|\ln B| \gtrsim 2$ constitutes moderate-to-strong evidence against the more complex model, confirming that the introduction of the features is statistically unjustified by the data. We observe that $\ln B$ is slightly higher for the PSC oscillatory feature for both the \texttt{CamSpec} likelihoods, but it is still disfavoured with respect to the $\Lambda$CDM model. 
This result can be interpreted as a combination of two penalization effects: on one hand, the evidence is suppressed by a severe and often arbitrary Occam's razor penalty driven by the prior volume assigned to the amplitude parameter $A_X$; on the other hand, it is further suppressed by the prior range scanned in the other additional parameters, e.g. the frequency and the phase, accounting for a well-justified penalty due to the look-elsewhere effect. This combined volume suppression conceals the fact that locally we may observe a strongly favouring $\Delta\chi^2$ for a given frequency.

To disentangle these effects and properly quantify the statistical significance of these localized anomalies, we adopt a framework following Ref.~\cite{Bayer:2020pva} that maps local peaks to a calibrated global significance. This requires moving through three sequential quantities: the local fit quality ($q_L$), the local Bayesian mass ($q_B$), and the final look-elsewhere-corrected frequentist statistic ($q_S$).

The baseline for our local significance is the profile likelihood ratio test statistic, defined as
\begin{equation}
    q_L \equiv 2 \ln \sround{ \frac{ \sup_{\bm{\theta} \in \bm{\Theta}_1 }\mathcal{P}\round{\bm{x} \, | \,\bm{\theta}}}{ \sup_{\bm{\theta} \in \bm{\Theta}_0} \mathcal{P} \round{\bm{x}\, | \, \bm{\theta}}} } \; ,
\end{equation}
where $\mathcal{P} \round{\bm{x}\,|\,\bm{\theta}}$ is the likelihood function, while $\bm{\Theta}_0$ and $\bm{\Theta}_1$ represent the parameter space allowed under the null ($H_0$) and alternative ($H_1$) hypotheses, respectively.\footnote{Notice that, when computing the likelihood ratio, $\sup_{\bm{\theta} \in \bm{\Theta}_0} \mathcal{P} \round{\bm{x}\, | \, \bm{\theta}}$ is not necessarily the value that maximizes the likelihood for the sampled data} 
Under our assumption of flat priors, maximizing the likelihood to compute $q_L$ is equivalent to identifying the MAP log-posterior density. Numerically, for a Gaussian likelihood, $q_L$ corresponds to the local $-\Delta\chi^2$ (where the sign depends on our conventions), where $\sqrt{q_L}$ would yield the true statistical significance only in the absence of a look-elsewhere effect (LEE).

Because the primordial feature model allows for a wide search space in frequency and phase, random noise fluctuations generate a highly multi-modal posterior distribution. In this regime, selecting an appropriate \textit{point estimator} is crucial. While the standard MAP estimator identifies the peak of the posterior density, it lacks information about the volume of the parameter space supporting that mode. To address this, we consider the maximum posterior mass (MPM) estimator. Although both MAP and MPM are point estimators, the MPM framework evaluates the integrated posterior mass surrounding a peak rather than just its localized maximum density. In our context, since the analysis is performed locally around each identified peak, the primary advantage of adopting the MPM framework is not merely finding a more globally representative point; rather, its true utility lies in its mathematical structure, which intrinsically tracks the local posterior volume. 
Under the Laplace approximation, the local mass $m_i$ of the $i$-th peak is proportional to its posterior volume
\begin{equation}
    m_i \round{\bm{x}} = \mathcal{P} \round{ \bm{x} \,|\, \bm{\theta}_i}  \mathcal{P} \round{ \bm{\theta}_i } \round{2\pi}^{M/2} \sqrt{\det \bm{\Sigma}_i} \equiv \mathcal{P} \round{ \bm{x} \,|\, \bm{\theta}_i}  \mathcal{P} \round{ \bm{\theta}_i } V_\mathrm{post} \; ,
\end{equation}
where $\bm{\Sigma}_i$ is the local covariance matrix. 
The total global evidence of the extended model could then be computed as the sum over all modes, $\mathcal{Z} = \sum_i m_i$. To isolate the impact of a single localized anomaly and complete the analogy with the local likelihood ratio, we define the local Bayesian test statistic $q_B \equiv 2\ln B_{\text{local}}$ for a given dominant peak. Under our assumption of flat priors, this quantity naturally introduces the geometric Occam's razor penalty:
\begin{equation} \label{eqn:qB}
    q_B = q_L - 2 \ln \round{\frac{V_\text{prior}}{V_\text{posterior}}} \; .
\end{equation}
While the MPM is the point estimator that maximizes this $q_B$ quantity, it remains strictly Bayesian and explicitly depends on the full extra-parameter space, including the arbitrarily wide prior volume assigned to the amplitude parameter. As a result, $q_B$ lacks a direct frequentist significance interpretation, which makes it an intermediate tool rather than a final test statistic for the look-elsewhere effect.

To map these geometric volumes into a robust frequentist $p$-value, we introduce a trial factor $N$ to correct for the LEE, such that $P(Q_L > q_L) = N P_\text{local}(Q_L > q_L)$. Following the prescription of Ref.~\cite{Bayer:2020pva}, we exploit the relationship between the trial factor and the posterior volumes
\begin{equation} 
    N \equiv \frac{V^{\bcancel{A}}_\text{prior}}{V^{\bcancel{A}}_\text{posterior}} \; ,
\end{equation}
where the superscript $\bcancel{A}$ indicates that the volumes are computed exclusively over the non-amplitude parameters. This formulation utilises a specific reference prior that mathematically erases the arbitrary prior-dependence of the amplitude in~\cref{eqn:qB} effectively reducing $V_\text{posterior}$ to the $2\times2$ (for the LIN, LOG, and BUMP templates) and $3\times3$ (for the PSC templates) marginalised covariance matrix extracted from our posterior distributions. Finally, we construct a universal test statistic that, in analogy of what done with other hypotheses tests, we encode in the quantity 
\begin{equation}
    q_S \equiv q_L -2 \ln N + 2\ln \round{2 \pi q_L} - 2\ln t \; , 
    \label{eq:LEE_qs}
\end{equation}
where $t=1,2$ accounts for one or two-tailed tests, respectively. The final, look-elsewhere-corrected significance $S$ (in units of $\sigma$) is then given by
\begin{equation} \label{eq:lee_significance}
    S^2 \approx q_S - 2\ln \round{2 \pi q_S} +2\ln t \; .
\end{equation}

\begin{figure}
    \centering
    \includegraphics[width=0.98\linewidth]{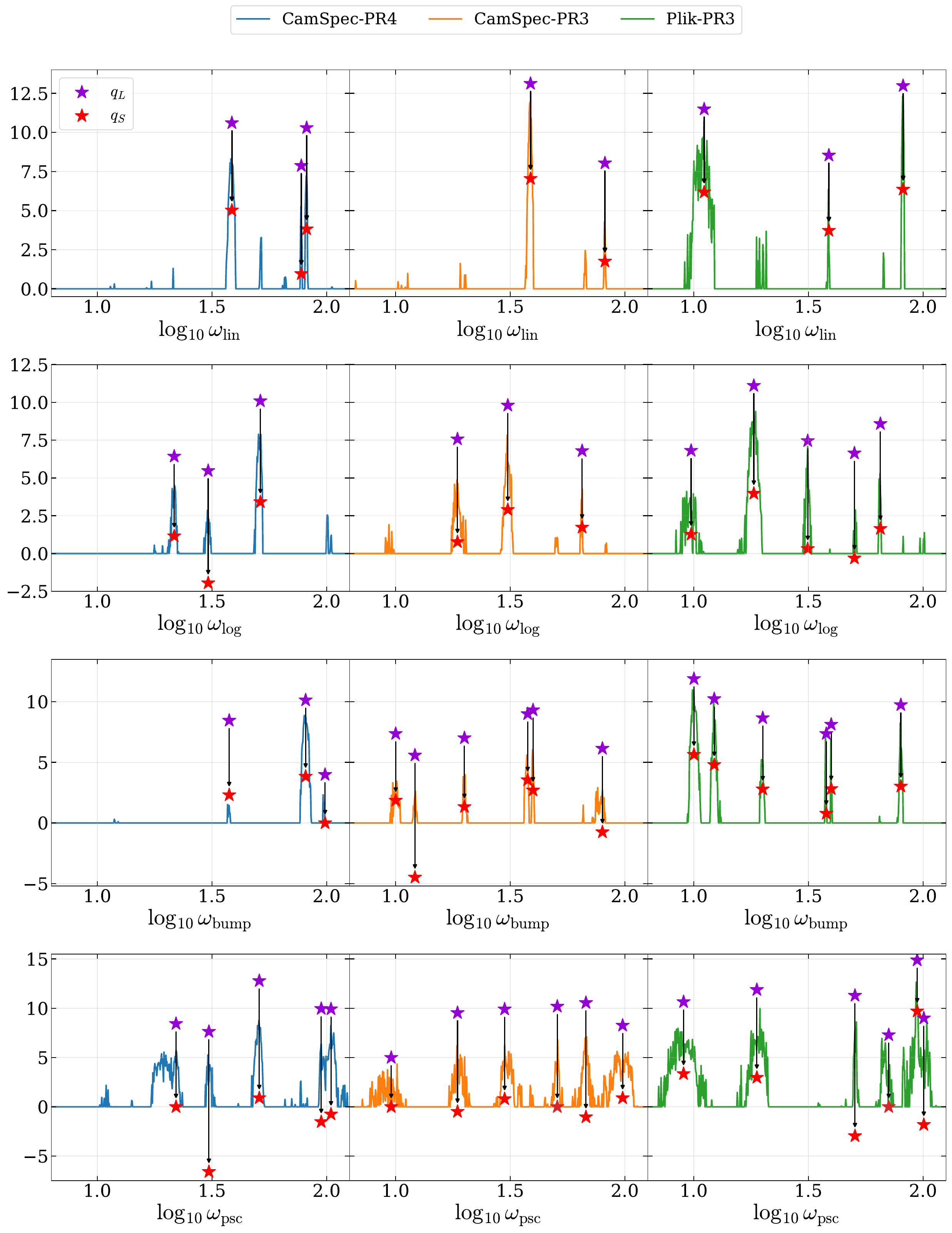}
    \caption{Results of the MPM analysis. Purple stars are the values obtained with the MAP analysis, that is $q_{L}=-\Delta\chi^2$; red stars are the $q_{S}$ from~\cref{eq:LEE_qs} for each minimum.}
    \label{fig:lee_correction}
\end{figure}

In~\cref{fig:lee_correction}, we show the values for the log likelihood ratio, $q_L$, and its LEE corrected version, $q_S$, for all the templates and the datasets considered in this study. As expected, it can be seen that the statistical significance of all the peaks is heavily affected by a corrected estimation of the LEE. 
The most significant detections for each template are: $2.1\sigma$ for the LIN template in \textit{CamSpec-PR3}, $1.5\sigma$ for the LOG template in \textit{Plik-PR3}, $1.9\sigma$ for the BUMP template in \textit{Plik-PR3}, $2.6\sigma$ for the PSC template in \textit{Plik-PR3}. 
We report the values of the significance for the best-fit values previously found in the MAP analysis in~\cref{tab:bestfits}. We notice that the best-fit frequency we found in the MAP analysis for the BUMP feature with \textit{CamSpec-PR3} does not correspond to the highest significance. The highest significance for this combination is $1.4\sigma$ for $\log_{10}\omega_\mathrm{bump}=1.58$. The same applies to the PSC template with \textit{CamSpec-PR3}, where the previously found best-fit has negative $q_S$, while the highest significance for this combination is $0.8\sigma$ for $\log_{10}\omega_\mathrm{psc}=1.48$. 
It is worth noting that all these values come from the PR3 of the \textit{Planck} data products. As soon as we constrain the classification we have just done to only PR4 data only, we get the following: $1.7\sigma$ for the LIN template, $1.3\sigma$ for the LOG template, $1.4\sigma$ for the BUMP template, and $0.74\sigma$ for the PSC template. It is important to stress that these significances are to be interpreted within the choice of the template, and are not to be considered as the global significance of a generic feature.

Finally, we remark that for the BUMP and PSC templates, the added structural complexity and parameter degeneracies introduced by the extra parameter, $\alpha$ and $k_r$, make it challenging to unambiguously identify the connected posterior region to be used for the volume evaluation. Due to these geometric ambiguities, the estimated global significances for these templates, particularly for the highest peaks, carry an inherent systematic uncertainty of approximately $0.5\sigma$.

\section{Forecasts for future CMB experiments} \label{sec:forecasts}
In this section we present forecasts for future CMB experiments using the same analysis pipeline adopted for the current-data analysis in the previous section. 
We generate two mock CMB datasets for a baseline featureless $\Lambda$CDM fiducial model with parameters fixed to $\omega_\text{c} = 0.1197$, $\omega_\text{b} = 0.02220$, $\tau_\text{reio} = 0.0577$, $100 \, \theta_\text{MC} = 1.04079$, $\ln\left(10^{10}A_\mathrm{s}\right) = 3.044$, and $n_\mathrm{s} = 0.9626$, and a fiducial models with an injected primordial-feature signal with feature parameters corresponding to the \textit{CamSpec-PR4} best-fits reported in~\cref{tab:bestfits}. 
We first describe the forecasting methodology and experimental specifications, and then present the forecast results.

As analytically demonstrated in~\cref{app:analytic}, polarisation spectra are not affected by the Sachs-Wolfe plateau and preserve sharper oscillatory features compared to temperature data, making future  polarisation measurements crucial for constraining these models.

\subsection{Forecast methodology}
We consider temperature ($T$), $E$-mode polarisation ($E$), and CMB lensing ($\phi$), but do not include $B$-modes, which are not expected to improve constraints for these primordial feature models. The harmonic coefficients of the observed fields are written as
\begin{equation}
    a_{\ell m}^X = s_{\ell m}^X + n_{\ell m}^X \, ,
\end{equation}
assuming the CMB signal $s_{\ell m}$ and the instrumental noise $n_{\ell m}$ decorrelated. Here $X \in \{T,E,\phi\}$, where $\phi$ denotes the lensing potential. The angular power spectrum of the observed coefficients takes the form
\begin{equation}
    \langle a_{\ell m}^{X*} a_{\ell' m'}^Y \rangle
    =
    \left(C_\ell^{XY} + \delta^{XY}\mathcal{N}_\ell^X\right)\delta_{\ell\ell'}\delta_{mm'} \, .
\end{equation}
Instrumental noise spectra for temperature and polarisation are modelled as isotropic white noise deconvolved with a Gaussian beam~\cite{Knox:1995dq}
\begin{equation}
    \mathcal{N}_\ell^X
    =
    \theta_{\rm FWHM}^2 \sigma_X^2
    \exp\!\left[\ell(\ell+1)\frac{\theta_{\rm FWHM}^2}{8\ln 2}\right] \, ,
    \label{eq:noise}
\end{equation}
where $\theta_{\rm FWHM}$ is the beam full width at half maximum and $\sigma_X$ is the map sensitivity. For lensing, we use the reconstruction noise $\mathcal{N}_\ell^\phi$ provided by the adopted experimental configuration.

We consider a joint SO plus LiteBIRD large-aperture telescope configuration. For LiteBIRD, we compute inverse noise weighted temperature and polarisation noise spectra from the channel sensitivities and beam sizes given in Ref.~\cite{LiteBIRD:2022cnt}, retaining only the seven channels in the $78$--$195\,{\rm GHz}$ range. For the SO, we use the publicly available goal-sensitivity noise curves for temperature, polarisation, and lensing~\cite{SimonsObservatory:2018koc}.\footnote{\url{https://github.com/simonsobs/so_noise_models}} In the combined setup, LiteBIRD provides large-scale temperature and $E$-mode information over $2 \le \ell \le 600$ with $f_{\rm sky}=0.7$, while the SO provides temperature and polarisation data over $601 \le \ell \le 3000$ on the common sky fraction $f_{\rm sky}=0.4$. 
Lensing reconstruction is taken exclusively from the SO over $30 \le \ell \le 3000$ with $f_{\rm sky}=0.4$.

The mock likelihood is constructed from the empirical power spectra in the standard full-sky approximation. For Gaussian CMB fields this corresponds to the exact Wishart form, which in practice we evaluate with the usual $f_{\rm sky}$ rescaling. For the data vector $\mathbf{a}=\{a_{\ell m}^T,a_{\ell m}^E,a_{\ell m}^\phi\}$ and a parameter set $\bm{\theta}$, the likelihood can be written as
\begin{equation}
    -2\ln\mathcal{L}
    =
    \sum_\ell (2\ell+1)f_{\rm sky}
    \left(
        \frac{D_\ell}{|\bar{C}_\ell|}
        +
        \ln\frac{|\bar{C}_\ell|}{|\hat{C}_\ell|}
        -
        3
    \right) ,
    \label{eq:gauss_like}
\end{equation}
where $\hat{C}_\ell$ denotes the mock observed covariance matrix and $\bar{C}_\ell(\bm{\theta})$ the theoretical covariance matrix, including noise. In our case, with three fields $\{T,E,\phi\}$ and neglecting $E\phi$ correlations, the determinant is
\begin{equation}
    |\bar{C}_\ell|
    =
    \bar{C}_\ell^{TT}\bar{C}_\ell^{EE}\bar{C}_\ell^{\phi\phi}
    -
    (\bar{C}_\ell^{TE})^2\bar{C}_\ell^{\phi\phi}
    -
    (\bar{C}_\ell^{T\phi})^2\bar{C}_\ell^{EE} \, ,
\end{equation}
and similarly for $|\hat{C}_\ell|$. The quantity $D_\ell$ is defined as
\begin{align}
    D_\ell
    &=
    \hat{C}_\ell^{TT}\bar{C}_\ell^{EE}\bar{C}_\ell^{\phi\phi}
    +
    \bar{C}_\ell^{TT}\hat{C}_\ell^{EE}\bar{C}_\ell^{\phi\phi}
    +
    \bar{C}_\ell^{TT}\bar{C}_\ell^{EE}\hat{C}_\ell^{\phi\phi}
    \notag\\
    &\quad
    -
    \bar{C}_\ell^{TE}
    \left(
        \bar{C}_\ell^{TE}\hat{C}_\ell^{\phi\phi}
        +
        2\hat{C}_\ell^{TE}\bar{C}_\ell^{\phi\phi}
    \right)
    -
    \bar{C}_\ell^{T\phi}
    \left(
        \bar{C}_\ell^{T\phi}\hat{C}_\ell^{EE}
        +
        2\hat{C}_\ell^{T\phi}\bar{C}_\ell^{EE}
    \right) \,.
\end{align}

Given a mock dataset, we perform forecasts by sampling the likelihood of~\cref{eq:gauss_like} and determine confidence intervals by exploring the parameter space with \texttt{PolyChord} with \texttt{Cobaya}.

\subsection{Forecast results}
\begin{figure}
    \centering
    \includegraphics[scale=0.5]{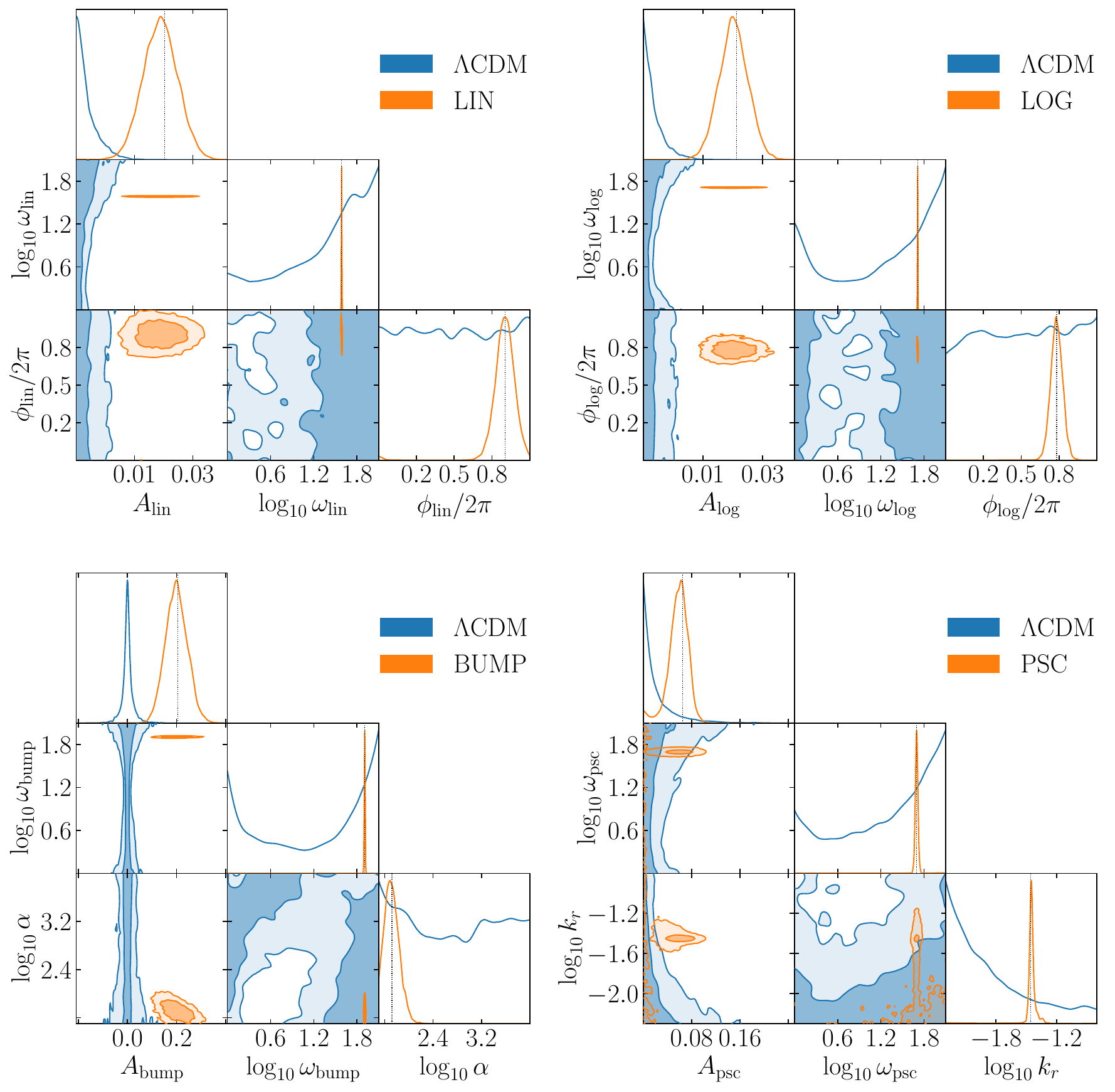}
    \caption{Marginalised joint 68\% and 95\% CL posterior distribution for feature parameters obtained using the LIN, LOG, BUMP, and PSC templates for the two different fiducial model (featureless PPS model in blue and \textit{CamSpec-PR4} best-fit feature template in orange). Grey dashed lines correspond to the parameter values used to generate the simulated inputs.}
    \label{fig:triangle_forecasts}
\end{figure}

We present the results of our forecast analysis for the four oscillatory templates in~\cref{fig:triangle_forecasts}, considering two distinct fiducial cosmologies: a \textit{null signal} case ($\Lambda$CDM with featureless PPS) and a \textit{feature detection} case (where the signal injected corresponds to the \textit{CamSpec-PR4} best-fits for each template).

In the absence of a primordial signal, future CMB observations will significantly tighten the current constraints on the feature amplitudes. For all templates (LIN, LOG, BUMP, and PSC), we find that SO+LiteBIRD will improve the 95\% CL upper limits to the level of $A_\mathrm{lin} < 0.009$, $A_\mathrm{log} < 0.008$, $-0.04 < A_\mathrm{bump} < 0.02$, and $A_\mathrm{psc} < 0.09$, representing an improvement of more than an order of magnitude over current \textit{Planck} constraints; see~\cref{tab:amplitude_forecast}.
These results are qualitatively consistent with previous forecast studies in the literature~\cite{Braglia:2022ftm}. In particular, the projected sensitivity to the feature amplitudes aligns well with the general expectations derived from information matrix techniques~\cite{Ballardini:2016hpi,Ballardini:2017qwq,Ballardini:2019tuc,Beutler:2019ojk,Euclid:2023shr}, with our full sampling-based analysis providing a robust validation for such estimates, which are typically based on the Gaussian approximation of the posterior.

In the scenario where a primordial feature is present in the data, our results demonstrate that the next generation of experiments will be able to characterise the signal with high precision. The amplitude of the features is well detected with a ratio $\sigma(A_X)/A_X$ of 0.20, 0.16, and 0.23 for the LIN, LOG, and BUMP templates, respectively. 
For these three templates, the additional feature parameters are also well constrained to $\log_{10}\omega_\mathrm{lin} = 1.637 \pm 0.002$ and $\phi_\mathrm{lin}/2\pi = 0.65 \pm 0.06$, $\log_{10}\omega_\mathrm{log} = 1.711 \pm 0.003$ and $\phi_\mathrm{log}/2\pi = 0.73 \pm 0.04$, $\log_{10}\omega_\mathrm{bump} = 1.91 \pm 0.01$ and $\log_{10} \alpha = 1.7 \pm 0.1$, all at 68\% CL.
For the PSC template, the injected signal is detected only at 68\% CL, while for the frequency and the clock mass we can only provide lower and upper bounds, respectively: $\log_{10}\omega_\mathrm{psc} > 1.7$ and $\log_{10} k_r < -1.4$ at 95\% CL.

Following the steps of our main analysis, we start by showing the Bayes factor with respect to the $\Lambda$CDM model (assumed as the fiducial model), which yields: $-4.9$ for the LIN template, $-5.1$ for the LOG template, $-3.6$ for the BUMP template, and $-0.8$ for the PSC template. As expected, $\Lambda\text{CDM}$ is preferred in all cases since it matches the true underlying fiducial cosmology. Notably, for the PSC template, the disfavour is significantly milder compared to the others. This behaviour indicates that a large fraction of the PSC prior volume maps into parameter configurations that are observationally equivalent to $\Lambda$CDM, when its envelope suppresses the oscillations. 

We can give an estimate of the LEE effect that affects the results of our forecasts. First of all let us stress that this estimate is possible only in the case in which we inject a signal in our fiducial model. 

When injecting a feature signal, the look-elsewhere-corrected global significances reach $3.7\sigma$ for the LIN template, $5.4\sigma$ for the LOG template, $4.0\sigma$ for the BUMP template, and $3.6\sigma$ for the PSC template. These high significance values reflect the enhanced sensitivity of the forecasted experimental setup. In particular, the significance of the LOG template ($5.4\sigma$) is a direct consequence of its frequency modulation; unlike the LIN template, whose oscillatory signal tends to be damped or out of phase at high multipoles due to projection effects, the LOG features retain a coherent high-frequency pattern in the high-$\ell$ regime; see~\cref{app:analytic}.

\begin{table}[t]
    \small
    \centering
    \renewcommand{\arraystretch}{1.4}
    \begin{tabular}{ccccc}
        \hline\hline
        \textbf{Fiducial model} & \boldmath $A_\mathrm{lin}$ & \boldmath $A_\mathrm{log}$ & \boldmath $A_\mathrm{bump}$ & \boldmath $A_\mathrm{psc}$ \\
        \hline
        Featureless & $< 0.0086$ & $< 0.0081$ & $0.000^{+0.018}_{-0.038}$ & $< 0.086$ \\
        Feature             & $0.0262 \pm 0.0052$ & $0.0227 \pm 0.0037$ & $0.149 \pm 0.034$ & $0.058^{+0.017}_{-0.012}\ (< 0.081)$ \\
    \hline\hline
    \end{tabular}
    \caption{Marginalised forecasted uncertainties on the amplitudes of the oscillatory feature templates for the two different simulated data. Uncertainties are quoted at 68\% CL while bounds at 95\% CL.}
    \label{tab:amplitude_forecast}
\end{table}

\section{Conclusions} \label{sec:conclusions}
The search for primordial oscillatory features in the cosmic microwave background provides a unique, yet challenging, window into the dynamics of the early Universe. In this work, we have systematically compared high-frequency oscillatory templates (LIN, LOG, BUMP, and PSC) with the most stringent unbinned \textit{Planck} likelihoods to date, contrasting the PR3 legacy release with the updated PR4 (\texttt{NPIPE}) data processing.  Although our selection does not cover the entire landscape of possible primordial feature models, this choice is physically justified: these benchmark templates capture the core phenomenological signatures while ensuring continuity and direct comparability with previous work in the literature. While our analysis identifies specific frequency bands (in the range $\omega \sim 10-100$) that formally yield an improved local fit to the data, achieving $\Delta\chi^2$ values up to $\sim -15$, a rigorous Bayesian evidence framework demonstrates that these improvements are insufficient to overcome the Occam penalty associated with the enlarged parameter space. 
Crucially, this conclusion is reinforced when accounting for the look-elsewhere effect through a calibrated frequentist test statistic ($q_S$), which accounts for the vast search volume of the extra parameters while avoiding the arbitrary amplitude prior dependence. After applying this correction, the statistical significance of the most prominent localised anomalies is drastically reduced, reaching a global significance of at most $1.7\sigma$ within the PR4 dataset. Furthermore, we observed that several marginal anomalies hinted at in earlier PR3-based studies, particularly at lower frequencies $\omega \lesssim 30$, are substantially mitigated when transitioning to the \texttt{CamSpec} likelihood based on \textit{Planck} PR4 products. 

We highlight analytically the fundamental physical limitation of current temperature-dominated datasets: the Sachs-Wolfe plateau and acoustic damping inherently blur high-frequency primordial signals. Conversely, $E$-mode polarisation transfer functions preserve these sharp oscillatory patterns with high contrast. Exploiting this physical distinction, our forecasts for a combined Simons Observatory and LiteBIRD configuration reveal the transformative potential of next-generation polarisation data. We project that future experiments will tighten the bounds on feature amplitudes by more than an order of magnitude. This leap in sensitivity will be essential not only for conclusively resolving the current tentative hints, but also for firmly probing the rich theoretical landscape of inflationary dynamics beyond the standard minimal $\Lambda$CDM framework.

\acknowledgments
We thank Erik Rosenberg for having provided us with foreground-subtracted and calibration-corrected binned spectra for the latest \texttt{CamSpec} analysis.
AR acknowledges support from ASI/INFN grant no. 2021-43-HH.0.
MB acknowledges financial support from the COSMOS network through the ASI (Italian Space Agency) Grants 2016-24-H.0, 201624-H.1-2018, 2020-9-HH.0 (participation in LiteBIRD phase A).
NB is supported by the Spanish grants PID2023-147306NB-I00 and CEX2023-001292-S (MCIU/AEI/10.13039/501100011033). NB is funded by the European Union (ERC, RELiCS, project number 101116027). Views and opinions expressed are however those of the author only and do not necessarily reflect those of the European Union or the European Research Council Executive Agency. Neither the European Union nor the granting authority can be held responsible for them. 
We acknowledge CINECA for the availability of high performance computing resources and support through the CINECA-INFN agreement.

\newpage
\appendix
\section{Results for the standard cosmological parameters}
\begin{figure}[h!]
    \centering
    \includegraphics[scale=0.5, trim={0 0 1.5cm 0}, clip]{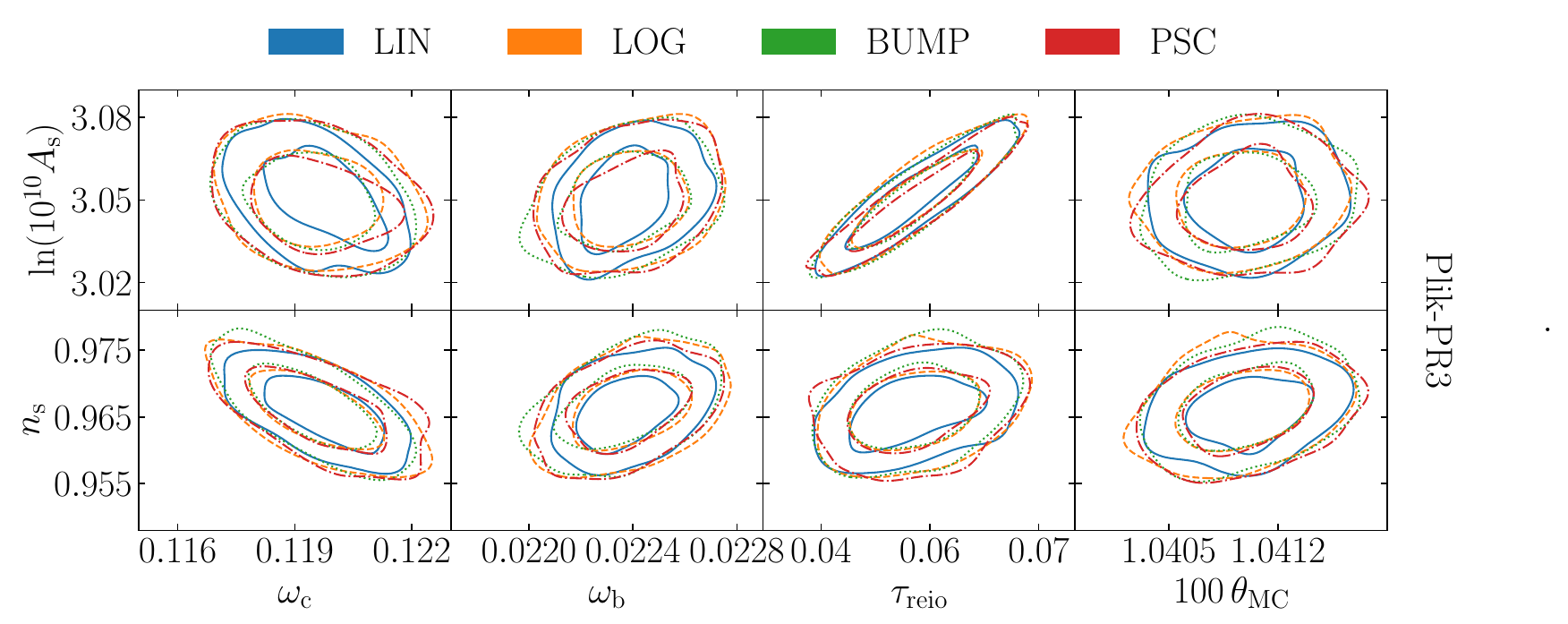}
    \includegraphics[scale=0.5, trim={0 0 1.5cm 0}, clip]{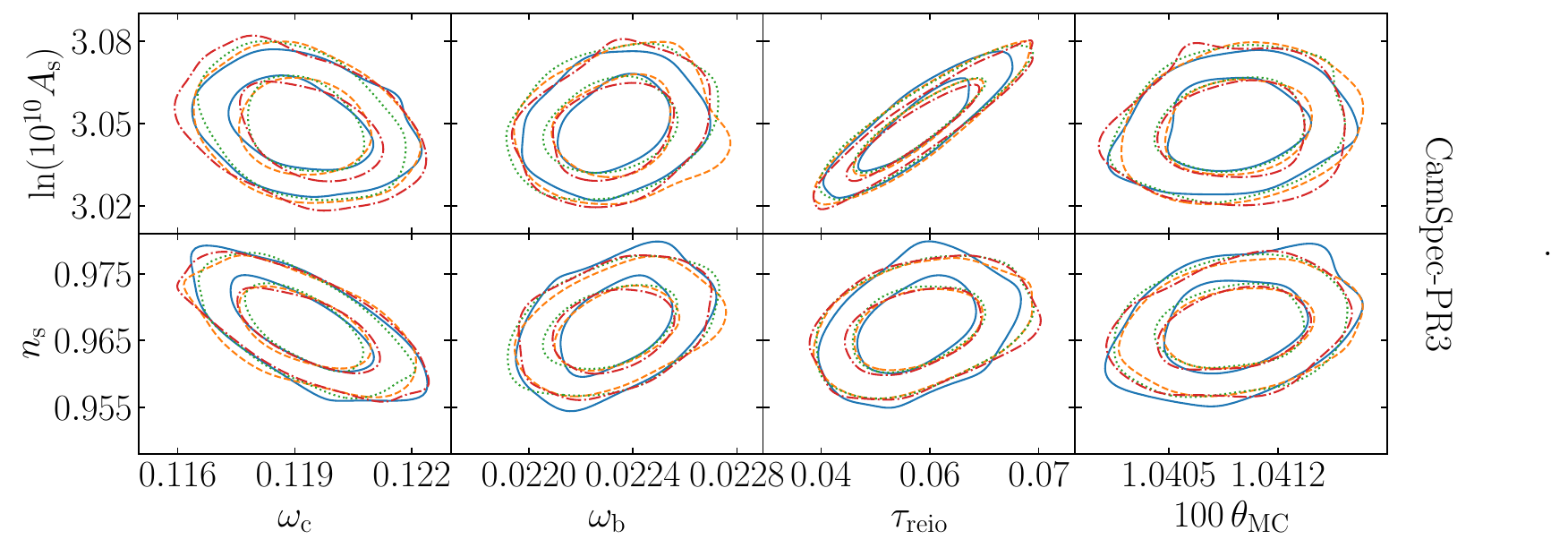}
    \includegraphics[scale=0.5, trim={0 0 1.5cm 0}, clip]{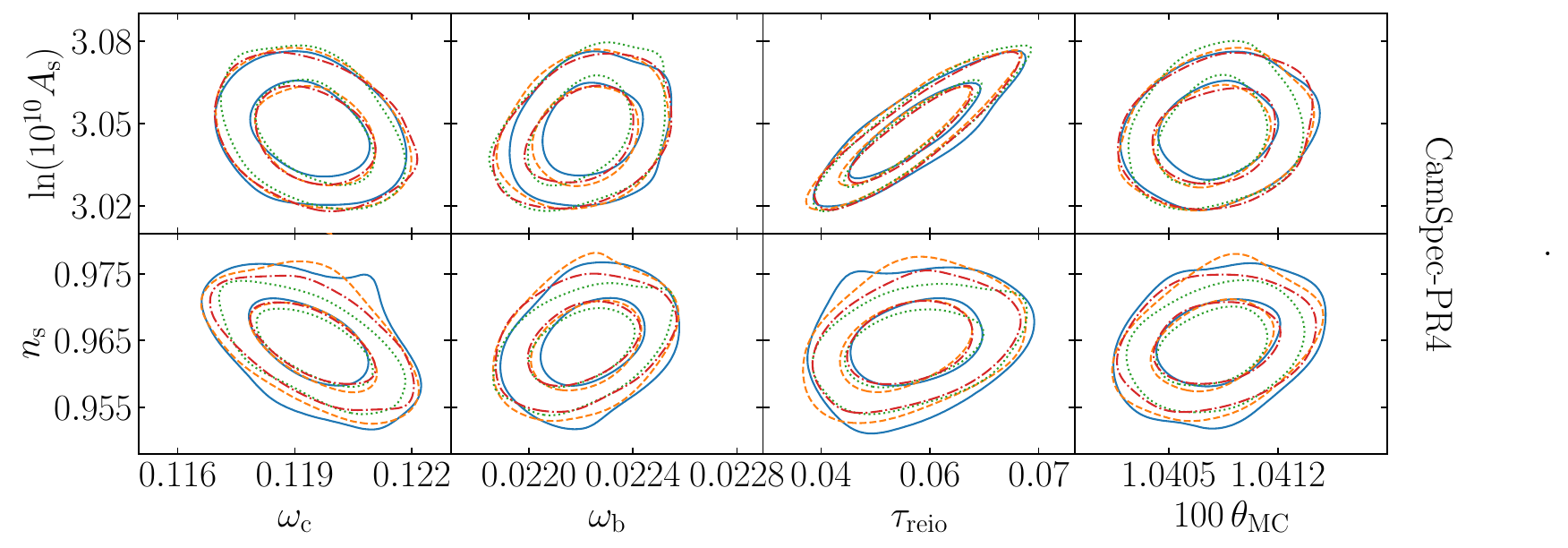}
    \caption{Marginalised joint 68\% and 95\% CL posterior distribution for standard cosmological parameters obtained using the LIN, LOG, BUMP, and PSC templates for the different datasets (\textit{Plik-PR3} upper panel, \textit{CamSpec-PR3} central panel, and \textit{CamSpec-PR4} bottom panel).}
    \label{fig:2D_standard}
\end{figure}

\label[appendix]{app:cosmo_params}
In this appendix, we collect the figures showing the marginalised two-dimensional posterior distributions, (\cref{fig:2D_standard}), together with the tables reporting the constraints on the cosmological parameters (\cref{tab:cosmo_params}) obtained for the three datasets analysed in this work. The constraints on the six standard cosmological parameters remain stable across the different parametrisations adopted for the PPS.

\begin{table}[h!]
    \small
    \centering
    \renewcommand{\arraystretch}{1.4} 
{\textit{Plik-PR3}}
\vskip 1mm 
\begin{tabular}{l c c c c}
\hline\hline
& \textbf{LIN} & \textbf{LOG} & \textbf{BUMP} & \textbf{PSC} \\
\hline
{\boldmath$\omega_\mathrm{c}$} & $0.1197^{+0.0012}_{-0.0009}$ & $0.1195\pm 0.0011$ & $0.1194\pm 0.0011$ & $0.1196\pm 0.0012$\\
{\boldmath$\omega_\mathrm{b}$} & $0.02237^{+0.00010}_{-0.00014}$ & $0.02239\pm 0.00015$ & $0.02237\pm 0.00016$ & $0.02238\pm 0.00015$\\
{\boldmath$\tau_\mathrm{reio}$} & $0.0575^{+0.0051}_{-0.0068}$ & $0.0582^{+0.0054}_{-0.0063}$ & $0.0578\pm 0.0059$ & $0.0580^{+0.0053}_{-0.0067}$\\
{\boldmath$100\theta_\mathrm{MC}$} & $1.04101\pm 0.00026$ & $1.04097\pm 0.00028$ & $1.04100^{+0.00027}_{-0.00030}$ & $1.04101\pm 0.00029$\\
{\boldmath$\ln(10^{10} A_\mathrm{s})$} & $3.050\pm 0.012$ & $3.051^{+0.011}_{-0.012}$ & $3.050\pm 0.012$ & $3.0490^{+0.009}_{-0.013}$\\
{\boldmath$n_\mathrm{s}$} & $0.9659\pm 0.0037$ & $0.9663\pm 0.0043$ & $0.9665\pm 0.0043$ & $0.9661\pm 0.0042$\\
\hline\hline
\end{tabular}

\vspace{0.5cm}

{\textit{CamSpec-PR3}}
\vskip 1mm 
\begin{tabular}{l c c c c}
\hline\hline
& \textbf{LIN} & \textbf{LOG} & \textbf{BUMP} & \textbf{PSC} \\
\hline
{\boldmath$\omega_\mathrm{c}$} & $0.1192^{+0.0011}_{-0.0013}$ & $0.1192\pm 0.0011$ & $0.1192\pm 0.0011$ & $0.1193^{+0.0013}_{-0.0011}$\\
{\boldmath$\omega_\mathrm{b}$} & $0.02233\pm 0.00015$ & $0.02234\pm 0.00016$ & $0.02232\pm 0.00017$ & $0.02233\pm 0.00015$\\
{\boldmath$\tau_\mathrm{reio}$} & $0.0583\pm 0.0053$ & $0.0583^{+0.0055}_{-0.0062}$ & $0.0583^{+0.0054}_{-0.0066}$ & $0.0581^{+0.0055}_{-0.0070}$\\
{\boldmath$100\theta_\mathrm{MC}$} & $1.04092\pm 0.00031$ & $1.04095\pm 0.00031$ & $1.04090\pm 0.00031$ & $1.04091\pm 0.00031$\\
{\boldmath$\ln(10^{10} A_\mathrm{s})$} & $3.049\pm 0.011$ & $3.049\pm 0.012$ & $3.049^{+0.011}_{-0.012}$ & $3.048^{+0.011}_{-0.013}$\\
{\boldmath$n_\mathrm{s}$} & $0.9674\pm 0.0051$ & $0.9669\pm 0.0042$ & $0.9671\pm 0.0044$ & $0.9667^{+0.0038}_{-0.0045}$\\
\hline\hline
\end{tabular}

\vspace{0.5cm}

{\textit{CamSpec-PR4}}
\vskip 1mm 
\begin{tabular}{l c c c c}
\hline\hline
& \textbf{LIN} & \textbf{LOG} & \textbf{BUMP} & \textbf{PSC} \\
\hline
{\boldmath$\omega_\mathrm{c}$} & $0.11938\pm 0.00098$ & $0.1195^{+0.0011}_{-0.0009}$ & $0.1195^{+0.0010}_{-0.0009}$ & $0.1195\pm 0.0010$\\
{\boldmath$\omega_\mathrm{b}$} & $0.02224\pm 0.00013$ & $0.02222\pm 0.00013$ & $0.02220\pm 0.00014$ & $0.02221\pm 0.00014$\\
{\boldmath$\tau_\mathrm{reio}$} & $0.0581\pm 0.0058$ & $0.0568^{+0.0053}_{-0.0063}$ & $0.0577^{+0.0054}_{-0.0068}$ & $0.0574^{+0.0053}_{-0.0060}$\\
{\boldmath$100\theta_\mathrm{MC}$} & $1.04082\pm 0.00025$ & $1.04079\pm 0.00025$ & $1.04079\pm 0.00023$ & $1.04080\pm 0.00026$\\
{\boldmath$\ln(10^{10} A_\mathrm{s})$} & $3.048\pm 0.011$ & $3.046^{+0.011}_{-0.013}$ & $3.047^{+0.011}_{-0.014}$ & $3.046^{+0.011}_{-0.012}$\\
{\boldmath$n_\mathrm{s}$} & $0.9646^{+0.0043}_{-0.0039}$ & $0.9647^{+0.0039}_{-0.0049}$ & $0.9640\pm 0.0039$ & $0.9647\pm 0.0042$\\
\hline\hline
\end{tabular}
\caption{Mean values and marginalised constraints at 68\% CL on the standard cosmological parameters for the
different datasets.}
\label{tab:cosmo_params}
\end{table}

\section{Best-fits feature parameters} \label[appendix]{app:bestfits}
In this appendix, we collect the tables reporting all the local best-fit values of the feature parameters, with an improvement in terms of $\Delta \chi^2$ above a given threshold; see~\cref{tab:osc_params_1,tab:osc_params_2,tab:osc_params_3,tab:osc_params_4}. These complement the global best-fits presented in~\cref{tab:bestfits}. 

\begin{table}[h!]
    \small
    \centering
    \renewcommand{\arraystretch}{1.4} 
    \begin{tabular}{ccccc}
    \hline\hline
    \textbf{Dataset}
    & \boldmath $A_\mathrm{lin}$ 
    & \boldmath $\log_{10}\omega_\mathrm{lin}$ 
    & \boldmath ${\phi_\mathrm{lin}}/2\pi$ 
    & \boldmath $\Delta\chi^2$ \\
    \hline
    & $0.014$ & 1.05 & $0.82$ & $-11.5$ \\[-4pt]
    \textit{Plik-PR3} & $0.021$ & 1.59 & $0.76$ & $-8.5$ \\[-4pt]
    & $0.035$ & 1.91 & $0.88$ & $-12.7$ \\
    \hline
    \multirow{2}{*}[2pt]{\textit{CamSpec-PR3}}
    & $0.027$ & 1.59 & $0.76$ & $-13.1$ \\[-4pt]
    & $0.029$ & 1.91 & $0.87$ & $-8.0$ \\
    \hline
    & $0.023$ & 1.58 & $0.83$ & $-10.6$ \\[-4pt]
    \textit{CamSpec-PR4} & $0.024$ & 1.89 & $0.21$ & $-7.9$ \\[-4pt]
    & $0.029$ & 1.91 & $0.90$ & $-10.3$ \\
    \hline\hline
    \end{tabular}
    \caption{Best-fit values of the template parameters. We also report $\Delta \chi^2 \equiv \chi^2_\mathrm{lin} - \chi^2_\mathrm{\Lambda CDM}$. Negative values of $\Delta \chi^2$ correspond to a better fit of the featureless PPS.}
    \label{tab:osc_params_1}
\end{table}

\begin{table}[h!]
    \small
    \centering
    \renewcommand{\arraystretch}{1.4} 
    \begin{tabular}{ccccc}
    \hline\hline
    \textbf{Dataset} 
    & \boldmath $A_\mathrm{log}$ 
    & \boldmath $\log_{10}\omega_\mathrm{log}$ 
    & \boldmath ${\phi_\mathrm{log}}/2\pi$ 
    & \boldmath $\Delta\chi^2$ \\
    \hline
    & $0.009$ & 0.99 & $0.70$ & $-6.8$ \\[-4pt]
    & $0.015$ & 1.26 & $0.31$ & $-11.1$ \\[-4pt]
    \textit{Plik-PR3} & $0.016$ & 1.50 & $0.93$ & $-7.5$ \\[-4pt]
    & $0.021$ & 1.70 & $0.74$  & $-6.6$ \\[-4pt]
    & $0.028$ & 1.81 & $0.69$ & $-8.6$ \\
    \hline
    & $0.014$ & 1.27 & $0.26$ & $-7.6$\\[-4pt]
    \textit{CamSpec-PR3} & $0.019$ & 1.49 & $0.94$ & $-9.8$ \\[-4pt]
    & $0.027$ & 1.81 & $0.96$ & $-6.8$ \\
    \hline
    & $0.011$ & 1.34 & $0.26$ & $-6.4$ \\[-4pt]
    \textit{CamSpec-PR4} & $0.013$ & 1.48 & $0.15$ & $-5.5$ \\[-4pt]
    & $0.024$ & 1.71 & $0.72$ & $-10.1$ \\
    \hline\hline
    \end{tabular}
    \caption{As in \cref{tab:osc_params_1}, but for the LOG template.}
    \label{tab:osc_params_2}
\end{table}

\begin{table}[h!]
    \small
    \centering
    \renewcommand{\arraystretch}{1.4} 
    \begin{tabular}{ccccc}
    \hline\hline
    \textbf{Dataset} 
    & \boldmath $A_\mathrm{bump}$ 
    & \boldmath $\log_{10}\omega_\mathrm{bump}$ 
    & \boldmath ${\log_{10}\alpha_\mathrm{bump}}$ 
    & \boldmath $\Delta\chi^2$ \\
    \hline
    & $0.017$ & 1.00 & 1.72 & $-11.9$ \\[-4pt]
    & $-0.053$ & 1.09 & 3.69 & $-10.2$ \\[-4pt]
    \multirow{2}{*}[2pt]{\textit{Plik-PR3}} & $-0.018$ & 1.30 & 2.44 & $-8.7$ \\[-4pt]
    & $0.032$ & 1.58 & 3.62 & $-7.4$ \\[-4pt]
    & $-0.047$ & 1.60 & 4.00 & $-8.1$ \\[-4pt]
    & $0.054$ & 1.90 & 2.71 & $-9.7$ \\
    \hline
    & $0.013$ & 1.00 & 1.78 & $-7.4$ \\[-4pt]
    & $-0.054$ & 1.08 & 3.95 & $-5.6$ \\[-4pt]
    \multirow{2}{*}[2pt]{\textit{CamSpec-PR3}} & $0.018$ & 1.30 & 2.77 & $-7.0$ \\[-4pt]
    & $0.030$ & 1.58 & 3.26 & $-9.0$ \\[-4pt]
    & $-0.052$ & 1.60 & 4.00 & $-9.3$ \\[-4pt]
    & $0.046$ & 1.90 & 2.65 & $-6.2$ \\
    \hline
    \multirow{2}{*}[2pt]{\textit{CamSpec-PR4}} & $0.024$ & 1.57 & 2.88 & $-8.5$ \\[-4pt]
    & $0.14$ & 1.90 & 1.78 & $-10.1$ \\
    \hline\hline
    \end{tabular}
    \caption{As in \cref{tab:osc_params_1}, but for the BUMP template.}
    \label{tab:osc_params_3}
\end{table}

\begin{table}[h!]
    \small
    \centering
    \renewcommand{\arraystretch}{1.4} 
    \begin{tabular}{cccccc}
    \hline\hline
    \textbf{Dataset} 
    & \boldmath $A_\mathrm{psc}$ 
    & \boldmath $\log_{10}\omega_\mathrm{psc}$ 
    & \boldmath $\log_{10}k_r$ 
    & \boldmath $\phi_\mathrm{psc}/2\pi$ 
    & \boldmath $\Delta\chi^2$ \\
    \hline
    & $0.034$ & 0.96 & $-1.50$ & $0.27$ & $-10.6$ \\[-4pt]
    & $0.042$ & 1.28 & $-1.49$ & $0.16$ & $-11.9$ \\[-4pt]
    \multirow{2}{*}[2pt]{\textit{Plik-PR3}} & $0.079$ & 1.70 & $-1.55$ & $0.98$ & $-11.3$ \\[-4pt]
    & $0.069$ & 1.85 & $-1.00$ & $0.82$ & $-7.3$ \\[-4pt]
    & $0.50$ & 1.97 & $-2.23$ & $0.24$ & $-14.9$ \\[-4pt]
    & $0.099$ & 2.00 & $-1.00$ & $0.0$ & $-9.0$ \\
    \hline
    & $0.027$ & 0.98  & $-1.50$ & $0.18$ & $-5.0$ \\[-4pt]
    & $0.042$ & 1.27 & $-1.48$ & $0.20$ & $-9.5$ \\[-4pt]
    \multirow{2}{*}[2pt]{\textit{CamSpec-PR3}} & $0.056$ & 1.48 & $-1.56$& $0.63$ & $-9.9$  \\[-4pt]
    & $0.077$ & 1.71 & $-1.55$ & $0.69$ & $-10.2$ \\[-4pt]
    & $0.076$ & 1.83 & $-1.17$& $0.17$ & $-10.6$ \\[-4pt]
    & $0.011$& 1.99 &  $-1.00$&  $0.10$ & $-8.3$ \\
    \hline
    & $0.011$ & 1.34 & $-2.06$ & $0.81$ & $-8.4$ \\[-4pt]
    & $0.031$ & 1.49 & $-1.66$ & $0.16$ & $-7.6$ \\[-4pt]
    \textit{CamSpec-PR4} & $0.070$ & 1.70 & $-1.55$ & $0.85$ & $-12.8$ \\[-4pt]
    & $0.045$ & 1.98 & $-2.23$ & $0.98$ & $-10.0$ \\[-4pt]
    & $0.092$ & 2.02 & $-1.05$ &  $0.21$ & $-9.9$ \\
    \hline\hline
    \end{tabular}
    \caption{As in \cref{tab:osc_params_1}, but for the PSC template.}
    \label{tab:osc_params_4}
\end{table}

\section{Residuals for TE and EE}
\begin{figure}
    \centering
    \includegraphics[width=\linewidth]{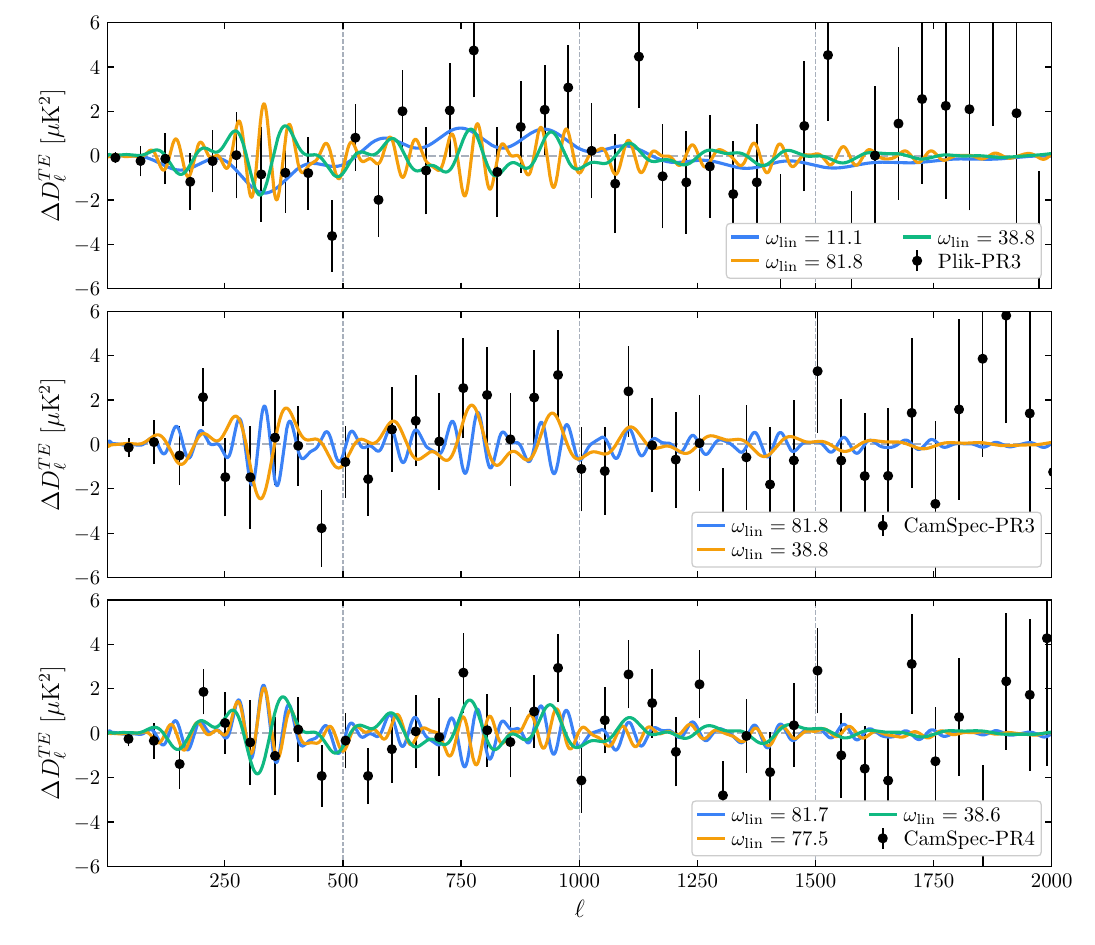}
    \caption{Residuals of the $TE$ angular power spectrum with respect to the best-fit featureless $\Lambda$CDM reference model for the LIN template. The black points with error bars show the residuals of the measured bandpowers relative to the corresponding binned $\Lambda$CDM prediction, while the coloured curves show the residuals of the best-fit LIN solutions with respect to the same reference. From top to bottom, the panels correspond to \textit{Plik-PR3}, \textit{CamSpec-PR3}, and \textit{CamSpec-PR4}. The labels in the legend indicate the frequencies of the best-fit solutions.}
    \label{fig:res_TE_LIN}
\end{figure}

\begin{figure}
    \centering
    \includegraphics[width=\linewidth]{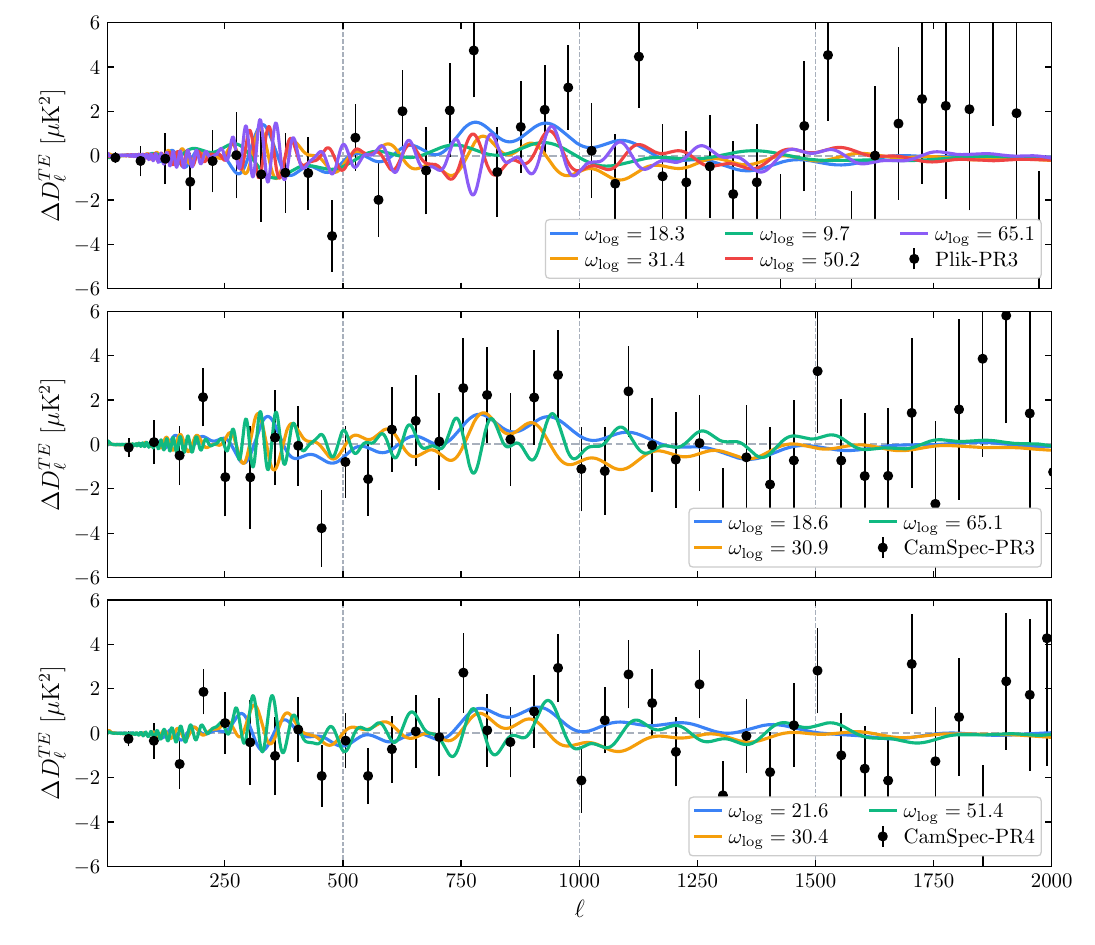}
    \caption{As in~\cref{fig:res_TE_LIN}, but for the LOG template.}
    \label{fig:res_TE_LOG}
\end{figure}

\begin{figure}[h!]
    \centering
    \includegraphics[width=\linewidth]{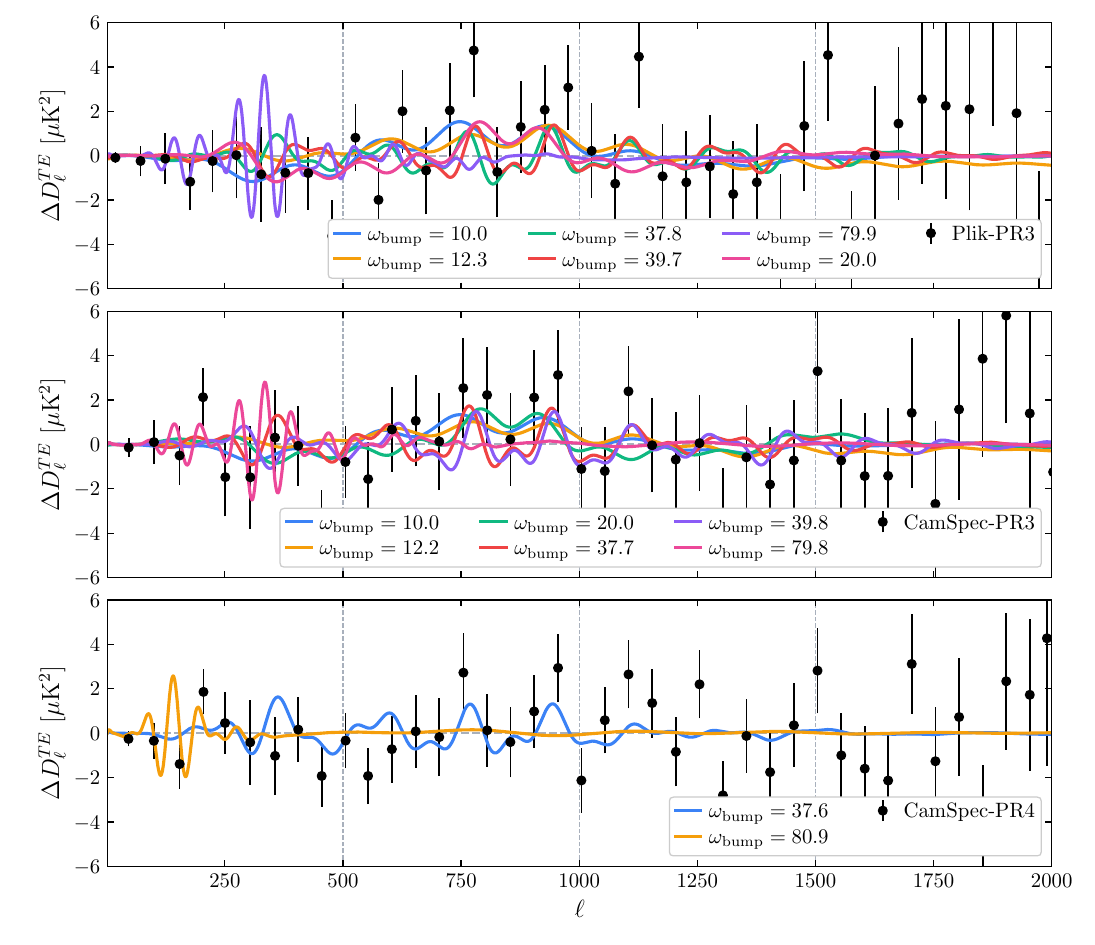}
    \caption{As in~\cref{fig:res_TE_LIN}, but for the BUMP template.}
    \label{fig:res_TE_BUMP}
\end{figure}

\begin{figure}[h!]
    \centering
    \includegraphics[width=\linewidth]{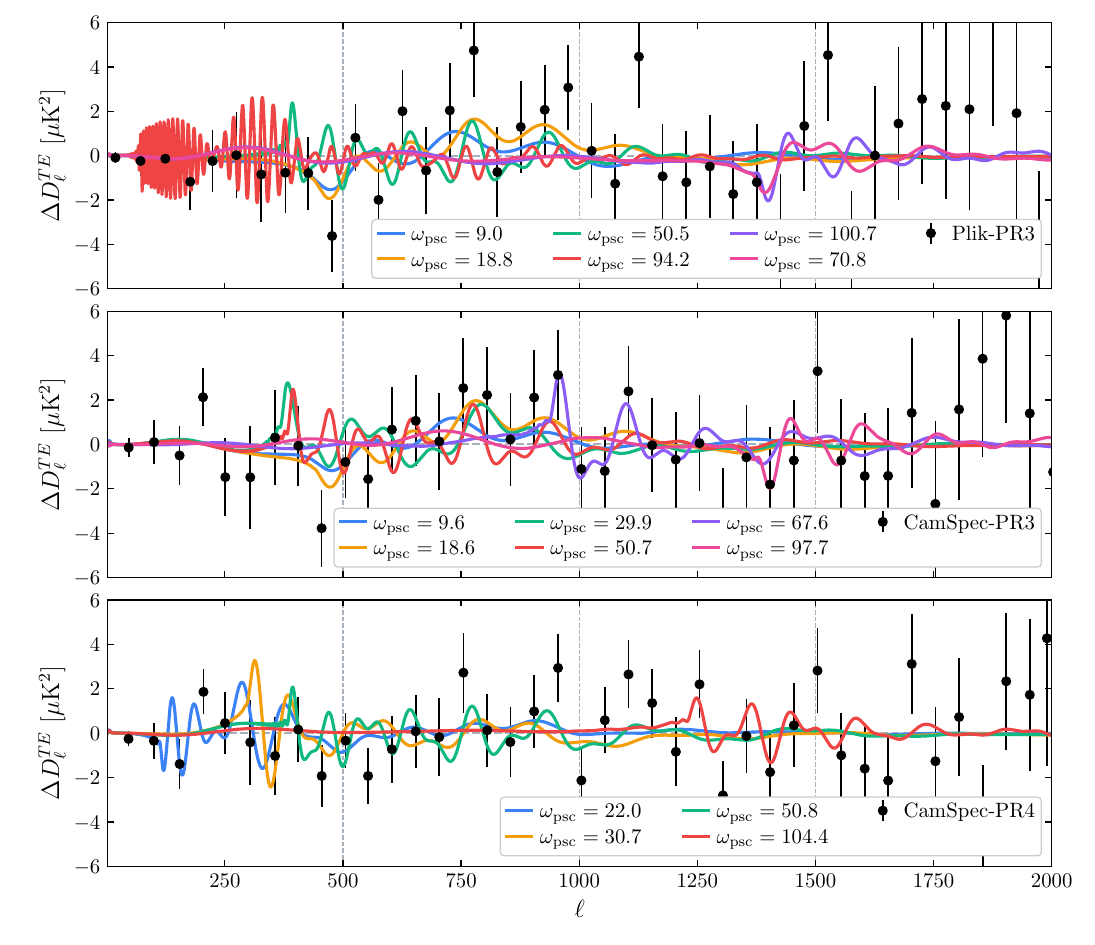}
    \caption{As in~\cref{fig:res_TE_LIN}, but for the PSC template.}
    \label{fig:res_TE_CLOCK}
\end{figure}

\begin{figure}[h!]
    \centering
    \includegraphics[width=\linewidth]{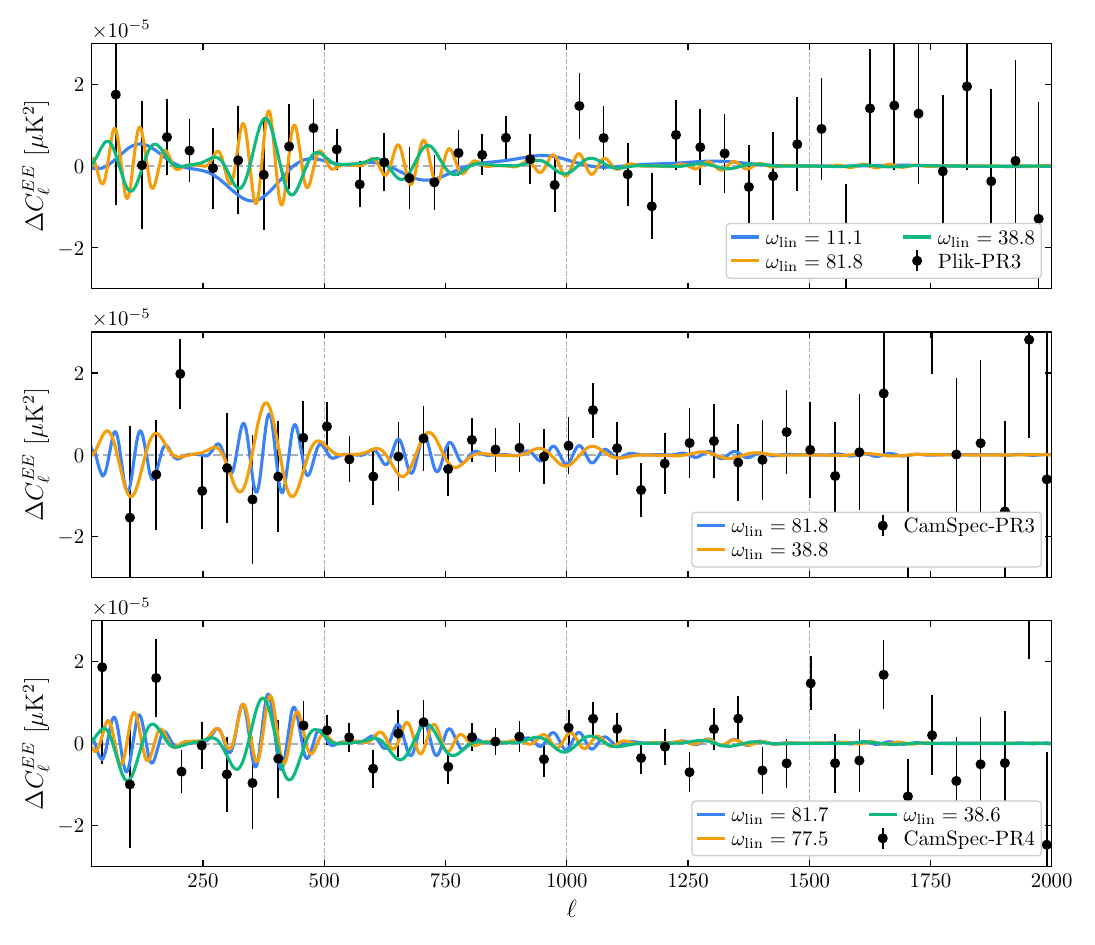}
    \caption{As in~\cref{fig:res_TE_LIN}, but for $EE$.}
    \label{fig:res_EE_LIN}
\end{figure}

\begin{figure}[h!]
    \centering
    \includegraphics[width=\linewidth]{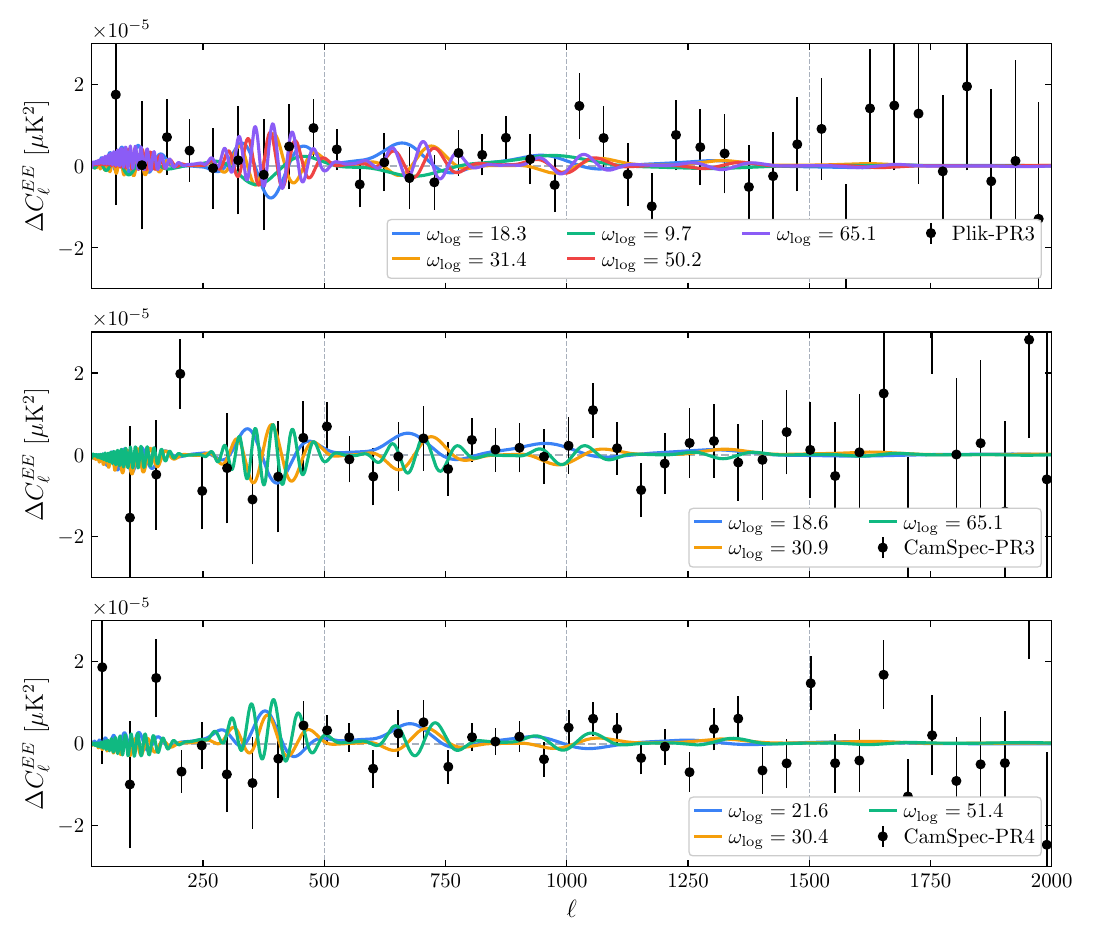}
    \caption{As in~\cref{fig:res_TE_LOG}, but for $EE$.}
    \label{fig:res_EE_LOG}
\end{figure}

\begin{figure}[h!]
    \centering
    \includegraphics[width=\linewidth]{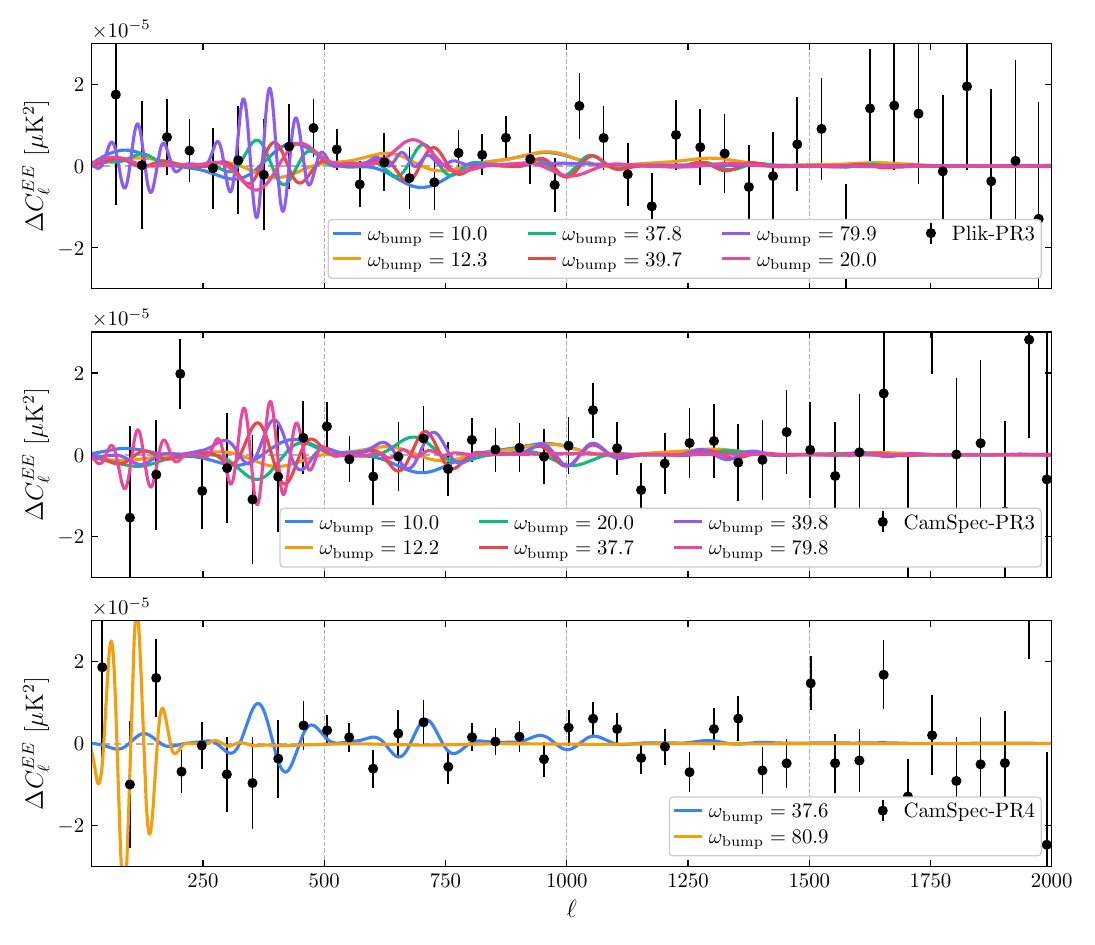}
    \caption{As in~\cref{fig:res_TE_BUMP}, but for $EE$.}
    \label{fig:res_EE_BUMP}
\end{figure}

\begin{figure}[h!]
    \centering
    \includegraphics[width=\linewidth]{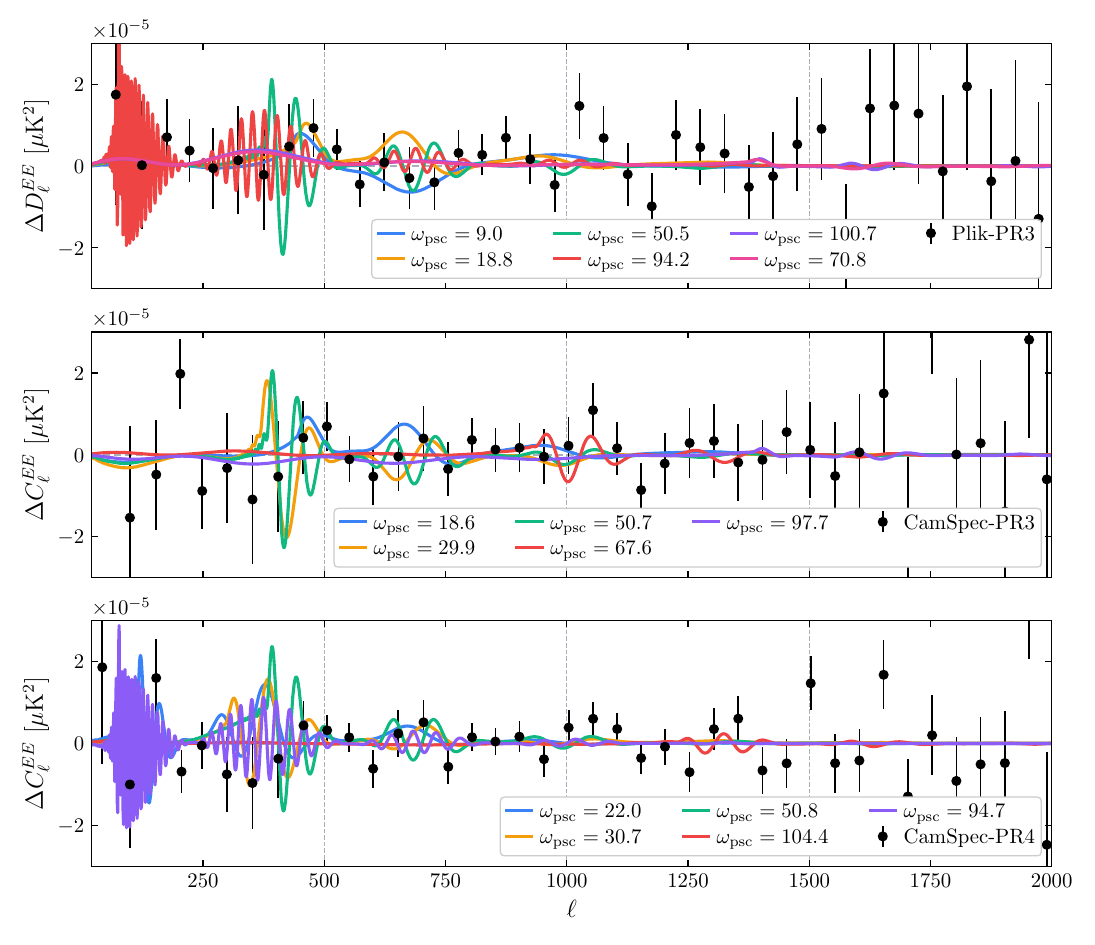}
    \caption{As in~\cref{fig:res_TE_CLOCK}, but for $EE$.}
    \label{fig:res_EE_CLOCK}
\end{figure}

\label[appendix]{app:residuals}
In this appendix, we show the residuals of the binned datasets, with $\Delta \ell = 50$, for the $TE$ angular power spectra in~\cref{fig:res_TE_LIN,fig:res_TE_LOG,fig:res_TE_BUMP,fig:res_TE_CLOCK} and for the $EE$ angular power spectra in~\cref{fig:res_EE_LIN,fig:res_EE_LOG,fig:res_EE_BUMP,fig:res_EE_CLOCK}. The different lines represent the best-fits for different frequencies corresponding to the parameters reported in \cref{app:bestfits}.

\section{Analytic considerations on the CMB power spectra} \label[appendix]{app:analytic}
We analyse the imprint of oscillatory features in the PPS on the CMB temperature and $E$-mode polarisation anisotropies; results are summarised in~\cref{fig:scaling_LIN,fig:scaling_LOG}.
We adopt the flat-sky approximation, where the celestial sphere is locally approximated by a plane perpendicular to the line of sight. 
In addition to flat-sky approximation, we assume instantaneous recombination occurring at the conformal time $\eta_*$, corresponding to a comoving distance $\chi_* = \eta_0 - \eta_*$. In this regime, scattering effects are negligible, and the transfer function is dominated by the intrinsic temperature fluctuation and the gravitational potential. By considering only the Sachs-Wolfe (SW) contribution, the transfer function is approximated as
\begin{equation} \label{eqn:SW_transfer}
    \Delta_\ell \approx \left[ \frac{1}{4}\delta_\gamma(k,\eta_*) + \Psi(k,\eta_*) \right] j_\ell(k\chi_*) \,,
\end{equation}
where $\delta_\gamma$ is the effective photon temperature fluctuation, $\Psi$ is the Newtonian potential defined with respect to the perturbed Robertson-Walker metric in longitudinal gauge 
\begin{equation} \label{eqn:RW_perturbed}
    \dd s^2 = a^2(\eta)\left[-\left(1+2\Psi\right)\dd \eta^2 + \left(1-2\Phi\right)g_{ij}\dd x^i\dd x^j\right] \,,
\end{equation}
and $j_\ell(x)$ is the spherical Bessel function of order $\ell$.

\subsection{The Sachs-Wolfe plateau}
We first consider the standard contribution from large scales (super-Hubble modes at decoupling). 
In this regime, photons have not had time to interact significantly or scatter. We can rewrite~\cref{eqn:SW_transfer} as
\begin{equation}
     \Delta_\ell(k) \approx \frac{1}{3}\Psi(k,\eta_*)j_\ell(k\chi_*)
\end{equation}
since during matter domination (MD) $\delta_\gamma/4 = 2\Phi/3$ and $\Psi = -\Phi$ assuming no anisotropic stress, $\Phi$ is the spatial curvature perturbation defined with respect to~\cref{eqn:RW_perturbed}.
The CMB temperature angular power spectrum results as 
\begin{equation}
    C_\ell \simeq \frac{2}{\pi} \int_0^{\infty} \dd k\, k^2 P_\Psi(k)\left[\frac{1}{3}j_\ell(k\chi_*)\right]^2 \,.
\end{equation}
In the MD epoch $\Psi = -3\zeta/5$ and assuming a scale-invariant PPS, i.e. $\PR(k) = 2\pi^2\As/k^3$, we obtain the standard result
\begin{equation} \label{eqn:SW_scaleinvariant}
    C_\ell \simeq \frac{4\pi}{25} \As \int_0^{\infty} \dd \ln{k}\, j_\ell(k\chi_*)^2 = \frac{\As}{25} \frac{2\pi}{\ell(\ell+1)} \,.
\end{equation}

In presence of a primordial oscillation, we need to rely on flat-sky approximation to perform the following integrals. 
In the flat-sky limit, for a thin shell at comoving distance $\chi_*$, the angular power spectrum is obtained by integrating out the line-of-sight (parallel) component of the wavevector $k_\parallel$ while fixing the transverse (perpendicular) component $k_\perp \simeq \ell / \chi_*$; the wavevector amplitude is $k = \sqrt{k_\perp^2 + k_\parallel^2}$.
Starting from the full-sky angular power spectrum, in cylindrical coordinates (aligned with the line of sight), it is $\dd^3 k = 2\pi k_\perp \dd k_\perp \dd k_\parallel$. For $\ell \gg 1$, the Bessel function is sharply peaked around $k\chi_* \simeq \ell$, meaning $k \simeq k_\perp$. We can approximate the square of the Bessel function as a Dirac delta in the transverse momentum as
\begin{equation}
    \frac{2}{\pi} j_\ell^2(k\chi_*) \propto \frac{1}{k^2 \chi_*^2} \delta\left(k_\perp - \frac{\ell}{\chi_*}\right) \,,
\end{equation}
this leads to
\begin{equation}
    C_\ell \simeq \frac{1}{\chi_*^2} \int_{-\infty}^{+\infty} \frac{\dd k_\parallel}{2\pi} \PR \left(\sqrt{k_\parallel^2 + k_\perp^2}\right) \bigg|_{k_\perp = \ell/\chi_*} \,.
\end{equation}
Assuming a scale-invariant PPS with a superimposed oscillation as 
\begin{equation}
    \PR(k) = \frac{2\pi^2}{k^3} \As \left[ 1 + A_X \sin\left(\omega_X\, \Xi_X(k) + \phi\right) \right] \,,
\end{equation}
the angular power spectrum splits into a smooth component, given by~\cref{eqn:SW_scaleinvariant}, and an oscillatory correction
\begin{equation}
    C_\ell = C_\ell^{\mathrm{smooth}} + \Delta C_\ell^{\mathrm{osc}} \,,
\end{equation}
where the correction term is given by
\begin{equation}
    \Delta C_\ell^{\mathrm{osc}} = \frac{\pi \As A_X}{25 \chi_*^2} \int_{-\infty}^{+\infty} \dd k_\parallel \frac{\sin\left[\omega_X\, \Xi_X\left(\sqrt{k_\perp^2 + k_\parallel^2}\right) + \phi\right]}{(k_\perp^2 + k_\parallel^2)^{3/2}} \,.
\end{equation}
The integral is dominated by modes near the point of closest approach, $k \simeq k_\perp$ (where $k_\parallel = 0$), where the phase of the oscillation varies most slowly. We can employ the stationary-phase approximation assuming $\omega_X\,  k_\perp \gg 1$ and approximate the denominator with $k^{-3} \simeq k_\perp^{-3}$.

For linear oscillations, we expand the wavenumber $k$ around $k_\parallel = 0$ up to second order, 
\begin{equation} \label{eqn:k_expansion}
    k = \sqrt{k_\perp^2 + k_\parallel^2} \simeq k_\perp + \frac{k_\parallel^2}{2k_\perp} \,,
\end{equation}
and the integral becomes
\begin{align}
    I_{\mathrm{osc}} &\simeq \frac{1}{k_\perp^3} \int_{-\infty}^{+\infty} \dd k_\parallel \sin\left(\tilde{\omega}_\mathrm{lin} k_\perp + \tilde{\omega}_\mathrm{lin} \frac{k_\parallel^2}{2k_\perp} + \phi\right) \simeq \frac{1}{k_\perp^3} \sqrt{\frac{2\pi k_\perp}{\tilde\omega_\mathrm{lin}}} \sin\left(\tilde\omega_\mathrm{lin} k_\perp + \phi + \frac{\pi}{4}\right)
\end{align}
where $\tilde\omega_\mathrm{lin} \equiv \omega_\mathrm{lin}/k_*$ and we have identified the standard Fresnel integral of the form $\int_{-\infty}^\infty \dd x\, e^{iax^2} = \sqrt{\pi/a}\,e^{i\pi/4}$.
Recalling $k_\perp = \ell/\chi_*$, we obtain
\begin{equation} \label{eqn:SW_lin}
    C_\ell \simeq \frac{2\pi \As}{25\ell^2} \left[1 + A_\mathrm{lin}\sqrt{\frac{\pi \ell_*}{2\omega_\mathrm{lin}\ell}} \sin\left(\omega_\mathrm{lin} \frac{\ell}{\ell_*} + \phi + \frac{\pi}{4}\right)\right] \,,
\end{equation}
where $\ell_* = k_* \chi_* \approx 695$.

For logarithmic oscillations the analogous of~\cref{eqn:k_expansion} is given by
\begin{equation}
    \ln\left(\frac{k_\perp}{k_*}\sqrt{1 + \frac{k_\parallel^2}{k_\perp^2}}\right) \simeq \ln \frac{k_\perp}{k_*} + \frac{k_\parallel^2}{2 k_\perp^2} 
\end{equation}
and the integral becomes
\begin{align}
    I_{\mathrm{osc}} &\simeq \frac{1}{k_\perp^3} \int_{-\infty}^{+\infty} \dd k_\parallel \sin\left[\omega_\mathrm{log} \ln \left(\frac{k_\perp}{k_*}\right) + \omega \frac{k_\parallel^2}{2k_\perp^2} + \phi\right] \\
    &\simeq \frac{1}{k_\perp^3} \left( k_\perp \sqrt{\frac{2\pi}{\omega_\mathrm{log}}} \right) \sin\left[\omega_\mathrm{log} \ln\left(\frac{k_\perp}{k_*}\right) + \phi + \frac{\pi}{4}\right] \,.
\end{align}
We obtain the explicit formula for the logarithmic case
\begin{equation} \label{eqn:SW_log}
    C_\ell \simeq \frac{2\pi \As}{25 \ell^2} \left\{ 1 + A_{\log} \sqrt{\frac{\pi}{2\omega_\mathrm{log}}} \sin\left[\omega_\mathrm{log} \ln \left(\frac{\ell}{\ell_*}\right) + \phi + \frac{\pi}{4}\right] \right\} \,.
\end{equation}

\begin{figure}[h!]
    \centering
    \includegraphics[width=\linewidth]{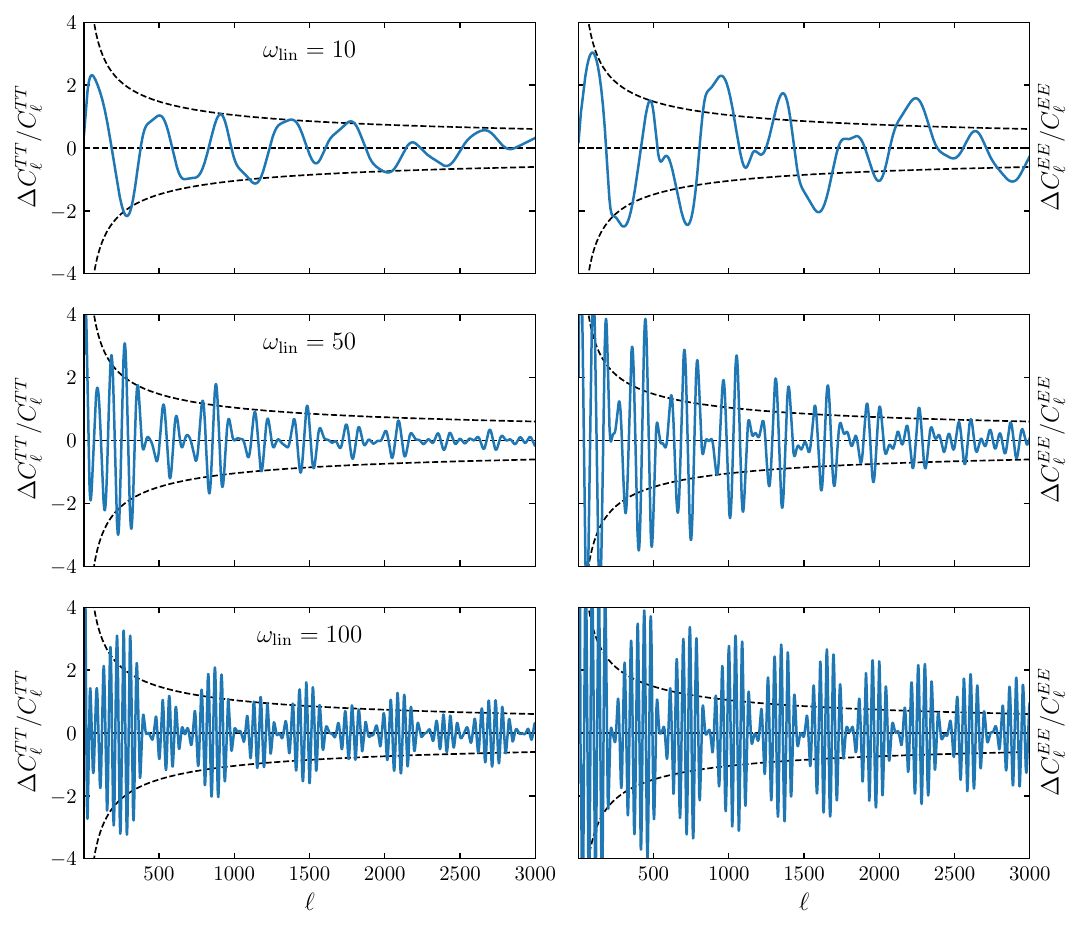}
    \caption{Fractional difference of the temperature (left panels) and $E$-mode polarisation power spectrum (right panels), scaled by $A_\mathrm{lin}/\sqrt{\omega_\mathrm{lin}}$, of the LIN template with $A_\mathrm{lin} = 0.05$ and $\phi_\mathrm{lin} = 0$ relative to a power-law PPS. In black-dashed lines, we show the expected scaling of the feature amplitude from \cref{eqn:SW_lin}, corresponding to $\sqrt{\pi k_* \chi_* /2\ell}$.}
    \label{fig:scaling_LIN}
\end{figure}

\Cref{eqn:SW_lin,eqn:SW_log} highlight the key effects of the two- to three-dimensional projection of the primordial oscillations, namely that the oscillation phase is shifted by $\pi/4$ and the feature amplitude is suppressed by a factor proportional to $\sqrt{\omega_X}$. A crucial distinction arises in the scaling with the multipole moment: while the amplitude of linear oscillations is further suppressed by a factor of $\sqrt{\ell}$~\cite{Adshead:2011jq,Huang:2012mr}, the logarithmic case remains unaffected~\cite{Huang:2012mr}; see~\cref{fig:scaling_LIN,fig:scaling_LOG}. This persistence occurs because the wavelength of logarithmic oscillations in $k$-space broadens proportionally to $k$, maintaining constant efficiency in the line-of-sight integration and preventing the features from being washed out at high multipoles.

It is evident from~\cref{eqn:SW_lin,eqn:SW_log} that linear oscillations imprint a pattern of constant frequency on the CMB angular power spectra, oscillating in multipole space $\ell$ with a fixed rate of $\omega_\mathrm{lin} / \ell_*$. Conversely, in the logarithmic case, the oscillations stretch out towards higher multipoles, exhibiting an effective local frequency that decays as $\omega_\mathrm{log} / \ell$.

\subsection{Sub-Hubble regime: acoustic oscillations}
On scales smaller than the sound horizon at recombination, i.e. $k\chi_* \gtrsim 1$, the evolution of photon perturbations is no longer dominated by gravity alone but is governed by the physics of the photon-baryon fluid. In this regime, we must account for the modulation introduced by the acoustic transfer functions. We adopt the tight-coupling approximation, where the photon-baryon fluid undergoes driven harmonic oscillations. The effective photon temperature fluctuation $\delta_\gamma/4$ is described by a damped oscillator driven by the Weyl potential $\Psi_W \equiv (\Phi+\Psi)/2$. Neglecting addition terms and considering the solution dominated by the cosine mode (adiabatic initial conditions), the transfer function is approximated as
\begin{equation}
    \Delta_\ell(k) \approx -\zeta(k)\cos(k \rs) \delta_\ell^{\mathrm{flat-sky}}(k) \,,
\end{equation}
where $\rs$ is the sound horizon at recombination and the projection kernel in the flat-sky limit is effectively the Bessel contribution. The angular power spectrum is given by the line-of-sight integral
\begin{equation}
    C_\ell = \frac{2}{\pi} \int \dd k\, k^2 \PR(k) \left[ \cos(k \rs) \delta_\ell^{\mathrm{flat-sky}}(k) \right]^2 \,.
\end{equation}
Expanding the cosine term via the identity $2\cos^2(x) = 1 + \cos(2x)$, we find that the acoustic physics modulates the signal by a factor $1/2$ (the average power) plus an oscillatory component $\cos(2k\rs)$ 
\begin{equation}
    \frac{A_X}{2} \sin\left(\omega \Xi(k) + \phi\right) + \frac{A_X}{2} \sin(\omega \Xi(k) + \phi)\cos(2 k \chi_*) \,.
\end{equation}
We apply the product-to-sum identity $\sin(A)\cos(B) = \frac{1}{2}[\sin(A+B) + \sin(A-B)]$ to the second term. This reveals the characteristic \textit{beating} of the signal into three components
\begin{equation}
    \frac{A_X}{2} \sin(\omega \Xi(k) + \phi)
    + \frac{A_X}{4} \sin\left(\omega \Xi(k) + 2\rs k + \phi\right)
    + \frac{A_X}{4} \sin\left(\omega \Xi(k) - 2\rs k + \phi\right) \,.
\end{equation}
The angular power spectrum is obtained by integrating over the line-of-sight wavevector $k_\parallel$ using the stationary-phase approximation as before. For linear oscillations the result is
\begin{align}
    \Delta C_\ell^\mathrm{osc} \simeq A_\mathrm{lin} \sqrt{\frac{\pi \ell_*}{2\omega_\mathrm{lin}\ell}} \Bigg\{ 
    &\sin\left(\omega_\mathrm{lin}\frac{\ell}{\ell_*} + \phi + \frac{\pi}{4}\right) \nonumber \\
    + &\frac{\sin\left[(\omega_\mathrm{lin} + 2k_*\rs)\frac{\ell}{\ell_*} + \phi + \frac{\pi}{4}\right]}{2\sqrt{|1 + 2k_*\rs/\omega_\mathrm{lin}|}} \nonumber \\
    + &\frac{\sin\left[(\omega_\mathrm{lin} - 2k_*\rs)\frac{\ell}{\ell_*} + \phi + \frac{\pi}{4}\right]}{2\sqrt{|1 - 2k_*\rs/\omega_\mathrm{lin}|}} 
    \Bigg\} \,.
\end{align}
We find two additional contributions to~\cref{eqn:SW_lin} where the sum mode $\omega_\mathrm{lin} + 2k_*\rs$ is damped more strongly than the difference mode $\omega_\mathrm{lin} - 2k_*\rs$ due to the different denominator. For low frequency $\omega_\mathrm{lin} \ll 2k_*\rs \simeq 14.4$ the first (pure) oscillatory term dominates.  
For high frequency $\omega_\mathrm{lin} \gg k_*\rs$ the last two term are equal $\omega_\mathrm{lin} \pm k_* \rs \simeq \omega_\mathrm{lin}$ and of the same order of the first term.

For logarithmic oscillations, we find
\begin{align}
    \Delta C_\ell^\mathrm{osc} \simeq A_\mathrm{log} \sqrt{\frac{\pi}{2\omega_\mathrm{log}}} \Bigg\{ 
    & \sin\left[\omega_\mathrm{log} \ln\left(\frac{\ell}{\ell_*}\right) + \phi + \frac{\pi}{4}\right] \notag\\
    + &\frac{\sin\left[\omega_\mathrm{log} \ln\left(\frac{\ell}{\ell_*}\right) + 2\frac{\rs}{\chi_*}\ell + \phi + \frac{\pi}{4}\right]}{2\sqrt{|1 \pm 2\rs \ell/(\chi_* \omega_\mathrm{log})|}}  \nonumber \\
    + &\frac{\sin\left[\omega_\mathrm{log} \ln\left(\frac{\ell}{\ell_*}\right) - 2\frac{\rs}{\chi_*}\ell + \phi + \frac{\pi}{4}\right]}{2\sqrt{|1 \mp 2\rs \ell/(\chi_* \omega_\mathrm{log})|}}  
    \Bigg\} \,.
\end{align}
Again, we find that logarithmic features do not suffer from the geometric $1/\sqrt{\ell}$ suppression seen in linear features because their oscillation period widens exactly as the projection kernel narrows. At low $\ell$, i.e. $\omega_\mathrm{log} \chi_*/\ell \gg 2\rs$, the logarithmic nature dominates with damping $\sim 1/\sqrt{\omega_\mathrm{log}}$ while at high $\ell$, i.e. $2\rs \gg \omega_\mathrm{log} \chi_*/\ell$, the acoustic linear oscillation dominates with damping $\sim 1/\sqrt{\ell}$.

\subsection{Diffusion and weak lensing dampings}
The finite mean free path of photons prior to recombination leads to the diffusion of photons out of overdensities, washing out perturbations on small scales including those imprinted by primordial oscillatory features. This is known as Silk damping.
We model this suppression by modifying the transfer function with an exponential damping factor
\begin{equation}
    \Delta_\ell(k) \to \Delta_\ell(k) e^{-(k/k_D)^2},
\end{equation}
where $k_D$ is the characteristic damping scale. For standard cosmological parameters, $k_D \approx 0.15 \, \mathrm{Mpc}^{-1}$, corresponding to $\ell_D \simeq k_D\chi_* \approx 2000$. For $\ell > \ell_D$, the primordial features become severely suppressed regardless the value of the amplitude $A_X$.

Finally there is a stronger damping starting at $\ell \sim 1000$. This is due to gravitational lensing. Without lensing the envelope follows the scaling derived above.

\subsection{E-mode polarisation transfer functions}
$E$-mode polarisation arises from the quadrupole anisotropy generated by Thomson scattering, which is sourced by the peculiar velocity of the photon fluid. Since the velocity field is $\pi/2$ out of phase with the density fluctuations that dominate the temperature spectrum, the acoustic peaks in $EE$ are anti-correlated with those in $TT$. 
The temperature spectrum is a complex superposition of the monopole (density), dipole (Doppler), and gravitational redshift (Sachs-Wolfe) effects. Conversely, the polarisation spectrum is devoid of the Sachs-Wolfe contribution, which dominates the temperature signal at large angular scales. 
In the temperature spectrum, the acoustic troughs are partially filled by the Doppler contribution, preventing the power from vanishing completely. In contrast, the polarisation oscillations are sharp and high-contrast, with minima that approach zero. 

\begin{figure}[h!]
    \centering
    \includegraphics[width=\linewidth]{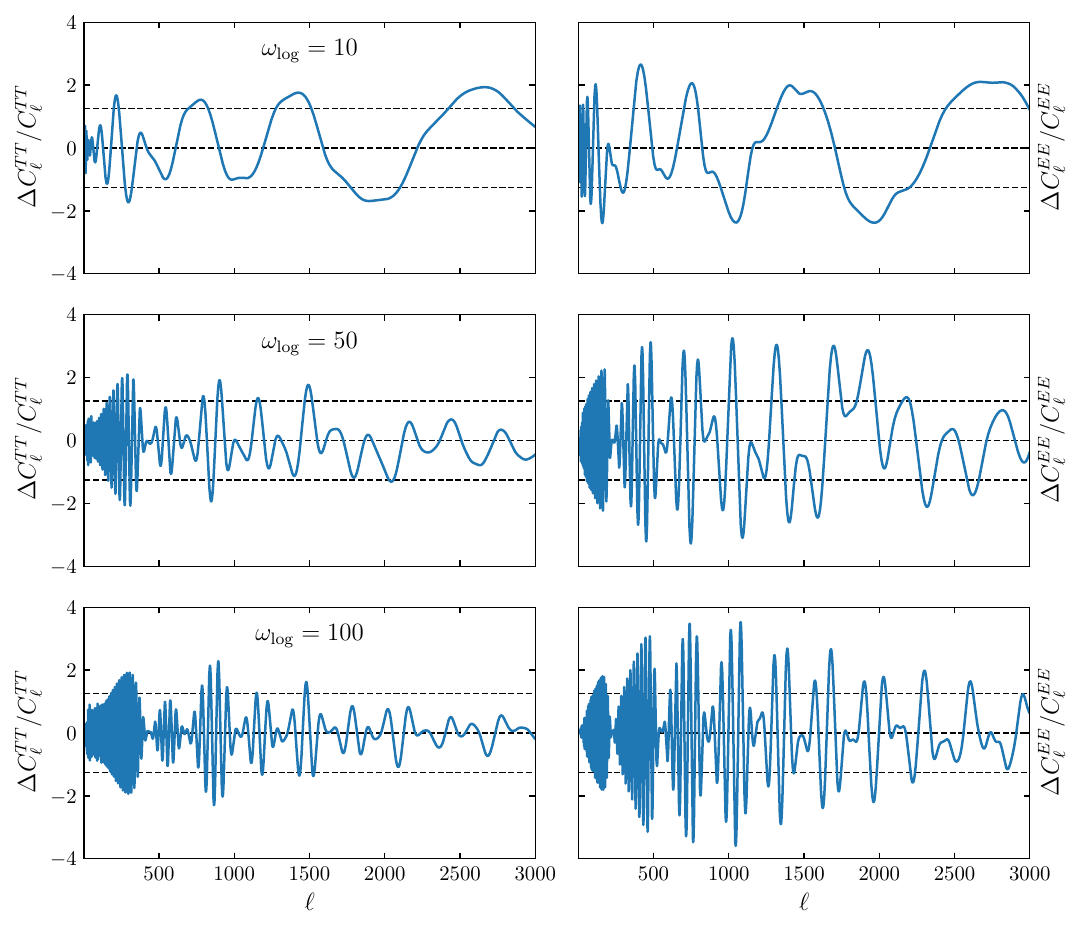}
    \caption{As in \cref{fig:scaling_LIN}, but for the LOG template. In black-dashed lines, we show the expected scaling of the feature amplitude from \cref{eqn:SW_log}, corresponding to $\sqrt{\pi /2}$.}
    \label{fig:scaling_LOG}
\end{figure}

\subsection{Signal-to-noise ratio}
To quantify the detectability of the oscillatory features from the CMB power spectra, we compute the cumulative signal-to-noise ratio as a function of the maximum multipole $\ell_\mathrm{max}$. 
For a given spectrum $X = TT, EE$, the cumulative signal-to-noise ratio is defined as
\begin{equation} \label{eqn:snr}
    \left(\frac{S}{N}\right)^2_X(\ell_{\max}) = \sum_{\ell=2}^{\ell_{\max}}
    \frac{\left(C_\ell^{X,\mathrm{model}} - C_\ell^{X,\mathrm{fid}}\right)^2}
    {\sigma^2\left(C_\ell^{X}\right)} \,,
\end{equation}
where $C_\ell^{X,\mathrm{fid}}$ denotes the fiducial $\Lambda$CDM power spectrum with power-law PPS.
The variance of the angular power spectrum is modelled assuming Gaussian statistics and is given by
\begin{equation}
    \sigma^2\left(C_\ell^{X}\right) = \frac{2}{(2\ell+1)\,f_{\mathrm{sky}}}\left(C_\ell^{X,\mathrm{fid}} + \mathcal{N}_\ell^{X}\right)^2 \,,
\end{equation}
where $f_{\mathrm{sky}}$ is the observed sky fraction and $\mathcal{N}_\ell^{X}$ denotes the instrumental noise power spectrum for the corresponding experiment.
This cumulative statistic allows us to identify the multipole ranges that contribute most significantly to the total signal-to-noise ratio and to compare the relative constraining power of the $TT$ and $EE$ spectra for the different feature templates considered. 

\begin{figure}[h!]
    \centering
    \includegraphics[width=\linewidth]{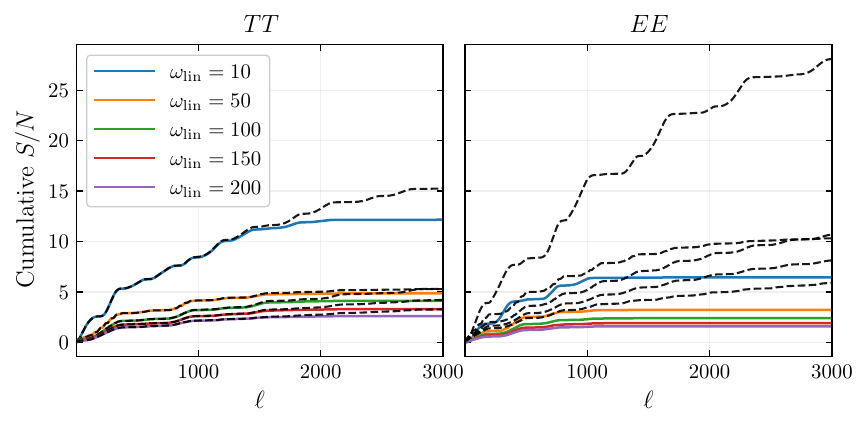}
    \caption{Cumulative signal-to-noise ratio defined according to \cref{eqn:snr} for the LIN template. Black dashed lines refers to the noiseless case.}
    \label{fig:snr_LIN}
\end{figure}

\begin{figure}[h!]
    \centering
    \includegraphics[width=\linewidth]{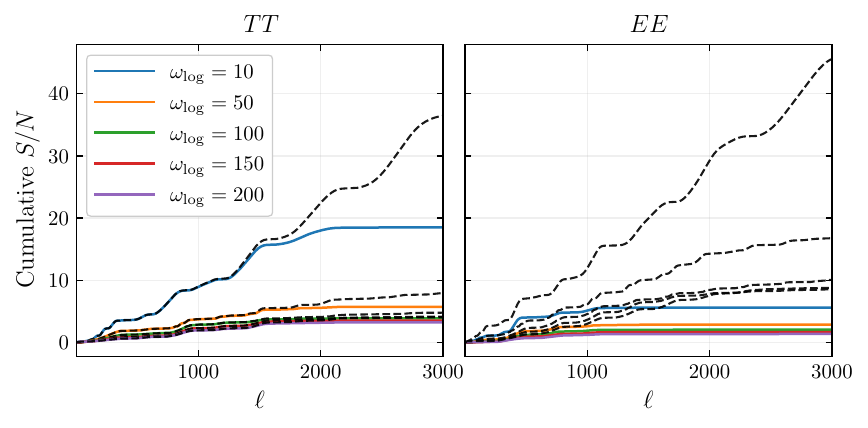}
    \caption{As for \cref{fig:snr_LIN}, but for the LOG template.}
    \label{fig:snr_LOG}
\end{figure}
\Cref{fig:snr_LIN,fig:snr_LOG} shows that the detectability of the oscillatory templates decreases as the oscillation frequency is increased,
but the degradation is gradual rather than abrupt. Most of the loss in cumulative signal-to-noise occurs when going from the lowest frequencies to intermediate values, while for $\omega \gtrsim 100$ the curves tend to cluster. This indicates a high-frequency regime in which the primordial oscillations are partially averaged by the CMB transfer functions and by the finite experimental resolution, so that further increasing the frequency produces only mild changes in the total signal-to-noise ratio.

The polarisation spectrum provides a larger cumulative signal-to-noise than temperature, especially at high multipoles. However, the main conclusion of this diagnostic is not the existence of a sharp frequency cutoff, but rather the presence of a broad transition to a regime of reduced sensitivity. This also explains why the amplitude constraints obtained in the likelihood analysis do not deteriorate dramatically over the frequency range considered.

For the LIN template, increasing $\omega_{\rm lin}$ produces faster oscillations in $k$ which are progressively smoothed by the projection onto 
angular power spectra. The resulting loss of sensitivity is therefore strongest at low-to-intermediate frequencies and then saturates. For the
LOG template, the effective period in multipole space is scale dependent, $\Delta \ell \sim 2\pi\ell/\omega_{\log}$, so that even relatively
large values of $\omega_{\log}$ can leave oscillations that are still resolved at high multipoles~\cref{fig:scaling_LOG}. This contributes to the absence of a sharp degradation of the constraints near the upper end of the frequency range used in the main analysis.

\bibliographystyle{JHEP}
\bibliography{bibliography}

\end{document}